\documentclass[a4paper,aps,preprintnumbers,amsmath,amssymb,
               superscriptaddress,floatfix,nofootinbib]{revtex4}
\usepackage{graphicx}
\graphicspath{{impact/}}
\usepackage{amsmath,amsfonts,amssymb}
\usepackage{mathrsfs}
\usepackage{color,xcolor}
\usepackage[normalem]{ulem} 
\newcommand{\rev}[1]{\textcolor{black}{#1}}
\usepackage{feynmp-auto}
\usepackage{slashed}
\usepackage{multirow}
\usepackage{hhline}
\usepackage{booktabs} 
\usepackage{float}    
\usepackage{subcaption}
\usepackage{tikz}
\usepackage{tikz-feynman}
\tikzfeynmanset{compat=1.1.0}

\begin{document}

\title{Odderon Form Factors in Reggeized Spin-2 Pomeron and Spin-3 Odderon Exchange in $pp$ and $p\bar p$ Elastic Scattering}

\author{Dominador F. Vaso, Jr.}
\email{dominadorjrvaso@gmail.com}
\affiliation{Department of Physics, Mindanao State University - Iligan Institute of Technology, Iligan City, 9200, Philippines}

\author{Prin Sawasdipol}
\email{p.namwongsa@kkumail.com}
\affiliation{Khon Kaen Particle Physics and Cosmology Theory Group (KKPaCT), Department of Physics, Faculty of Science, Khon Kaen University,123 Mitraphap Rd., Khon Kaen, 40002, Thailand}

\author{Jingle B. Magallanes}
\email{jingle.magallanes@g.msuiit.edu.ph}
\affiliation{Department of Physics, Mindanao State University - Iligan Institute of Technology, Iligan City, 9200, Philippines}

\author{Chakrit Pongkitivanichkul}
\email{chakpo@kku.ac.th}
\affiliation{Khon Kaen Particle Physics and Cosmology Theory Group (KKPaCT), Department of Physics, Faculty of Science, Khon Kaen University,123 Mitraphap Rd., Khon Kaen, 40002, Thailand}

\author{Daris Samart}
\email{darisa@kku.ac.th: corresponding author}
\affiliation{Khon Kaen Particle Physics and Cosmology Theory Group (KKPaCT), Department of Physics, Faculty of Science, Khon Kaen University,123 Mitraphap Rd., Khon Kaen, 40002, Thailand}

\date{\today}

\begin{abstract}
{\color{black}
We investigate the form-factor dependence of Reggeized tensor Pomeron and Odderon exchanges in high-energy elastic $pp$ and $p\bar p$ scattering. The spin structure is implemented through explicit covariant spin-2 and spin-3 projectors, kept factorized from the Reggeized scalar kernels, so that vertex effects can be separated from trajectory dynamics. Seven Odderon--proton form-factor parametrizations are tested against a global dataset including TOTEM $pp$ data at $\sqrt{s}=2.76$, $7$, $8$, and $13$~TeV and Tevatron $p\bar p$ data at $\sqrt{s}=1.80$ and $1.96$~TeV. A clear hierarchy is found. Six dipole, polynomial, Gaussian, and hybrid parametrizations give comparable fit qualities, $\chi^2_{\rm red}\simeq 1.44$--$1.48$, whereas a one-parameter exponential form,
$F_{\mathbb O}(t)=\exp[-B|t|/2]$, yields $\chi^2_{\rm red}=0.98$ for 138 degrees of freedom. The fitted couplings and Regge slopes remain comparatively stable across the form-factor choices, indicating that the improvement is driven mainly by the Odderon--proton vertex rather than by large compensating shifts in trajectory parameters. The exponential form admits an impact-parameter interpretation as a Gaussian transverse profile, with an effective radius $\sqrt{\langle b^2\rangle}=\sqrt{2B}\,\hbar c$. The extracted radii are of hadronic size and suggest a peripheral soft Odderon interaction. The shrinking $t$-range over which the single-Regge-exchange description remains accurate at increasing energy indicates the onset of absorptive and unitarity corrections. These results provide a compact phenomenological framework for connecting the $pp/p\bar p$ dip--bump difference with the transverse structure of $C$-odd color-singlet exchange.
}
\end{abstract}

\maketitle
\section{Introduction}\label{sec-1}

{\color{black}
Elastic proton--proton ($pp$) and proton--antiproton ($p\bar{p}$) scattering at high energies serves as a sensitive probe of the non-perturbative dynamics of the strong interaction \cite{antchev2019first}. Measurements of elastic differential cross sections over a wide range of momentum transfer squared, $t$, provide direct information on the spatial structure of hadrons and the nature of the exchanged objects carrying vacuum quantum numbers \cite{donnachie1984elastic, block2012forward}. Despite decades of intensive experimental and theoretical investigation, a complete description of elastic scattering within the framework of Quantum Chromodynamics (QCD) remains an open challenge, necessitating the continued application of effective and phenomenological approaches \cite{dremin2013elastic, bourrely2003impact}. Complementary holographic-QCD studies have also provided nonperturbative descriptions of high-energy elastic scattering, including Pomeron and Reggeon contributions to $pp$ and $p\bar p$ observables as well as Coulomb effects in both $pp/p\bar p$ and related pion--proton reactions \cite{Xie:2019soz,Watanabe:2019cvw,Liu:2022zsa,Zhang:2023nsk,Zhang:2024psj}. Prior to the modern understanding of QCD, the behavior of high-energy hadronic scattering was described using a phenomenological framework based on general principles such as the analyticity and unitarity of scattering amplitudes \cite{eden1966analytic}. This approach, known as Regge theory, describes $t$-channel exchanges as families of particles with increasing spin lying on Regge trajectories \cite{chew1961principle, gribov2003theory}. To account for the slow rise of total cross sections at asymptotically high energies \cite{froissart1961asymptotic, amendolia1973measurement}, the concept of the Pomeron was introduced; in modern QCD-inspired models, the Pomeron is interpreted as a color-singlet gluonic exchange \cite{low1975model, nussinov1975possible, Fadin:1975cb, Kuraev:1977fs}.

A central question in any modern treatment of high-energy hadronic scattering is the justification for using Reggeized exchanges rather than straightforward Feynman amplitudes built from fixed-spin fields. The necessity of Reggeization is dictated by three independent physical requirements that the elastic amplitude must satisfy, and which a fixed-spin treatment simultaneously violates. First, unitarity requires that the amplitude does not grow too rapidly; while a naive $t$-channel exchange of a fixed-spin-$J$ tensor field yields a growth of $\mathcal{A} \sim s^{J-1}$, Reggeization replaces the integer spin $J$ with a continuous trajectory $\alpha_X(t)$, restoring compatibility with the Froissart--Martin bound $\sigma_{\rm tot}(s) \leq C\ln^2(s/s_0)$. Second, the principles of analyticity and crossing require the amplitude to be analytic in $s$ and $t$ and to respect crossing symmetry between channels. Regge poles provide the unique analytic continuation of $t$-channel partial waves, encoding both the $s \to \infty$ asymptotic behavior and the spectrum of $t$-channel resonances within a single object. Third, the charge-conjugation structure of the $pp/p\bar{p}$ asymmetry, established with high precision by TOTEM and D\O~\cite{abazov2021odderon}, arises from the interference between $C=+1$ (Pomeron) and $C=-1$ (Odderon) exchanges. The relative phase between these contributions is not a free parameter but is strictly dictated by the Regge signature factors, $\frac{1}{2}[(-1)^{J_X} + e^{-i\pi\alpha_X(t)}]$. Without Reggeization, this phase remains undetermined, and the very observable that motivates the introduction of the Odderon cannot be predicted.

In Regge theory, the dominant $C=+1$ Pomeron exchange suggests that $pp$ and $p\bar{p}$ scattering should become identical at high energies \cite{pomeranchuk1958equality}. However, experimental studies reveal persistent, systematic differences, particularly in the dip--bump region \cite{breakstone1985measurement}. This necessitates a leading $C=-1$ exchange, the Odderon \cite{lukaszuk1973possible}, whose interference with the Pomeron naturally modifies the dip structure while leaving the total cross section largely unaffected \cite{antchev2019first, martynov2018did}. While joint analyses of TOTEM and D\O data provide \rev{strong experimental evidence for a $C$-odd exchange contribution} \cite{abazov2021odderon}, the exact functional dependence of the Odderon--proton vertex remains a significant source of theoretical uncertainty. Previous studies utilizing spin-2 Pomeron and spin-3 Odderon exchanges \cite{Ewerz:2016voh, ewerz2014model, magallanes2024odderon} have improved the description of differential cross sections, yet models often struggle to reproduce the data central values simultaneously across all energy scales \cite{csorgHo2021evidence}. This suggests that the underlying $t$-dependence, encoded in the hadronic form factors, may be more complex than the simple dipole or electromagnetic structures usually assumed \cite{sawasdipol2023effects}.

In this work, we extend the tensor Pomeron--Odderon framework by systematically investigating the role of hadronic form factors in elastic $pp$ and $p\bar{p}$ scattering in the Regge limit \cite{selyugin2015nucleon}. Our contributions are focused on four primary areas. We decouple the spin structure from the Regge propagator, separating the covariant tensor projector from the Reggeized scalar propagator to isolate form-factor dependence from trajectory parameters. This enables the first systematic comparison of seven Odderon--proton form-factor parametrizations within a coherent tensor-exchange framework. Furthermore, we \rev{test the degree of separation between vertex and trajectory dynamics: the trajectory parameters remain comparatively stable, while the fit quality varies significantly across form-factor choices. Finally, we identify an operational single-Regge-pole validity window, namely the $|t|$ range over which the Born-level Regge-pole description remains adequate. This window contracts as the center-of-mass energy increases, suggesting the progressive onset of absorptive and unitarity corrections at LHC energies.} The remainder of this paper is structured as follows. In Sec.~\ref{sec-2}, we establish the theoretical formalism and introduce the form-factor parameterizations. Sec.~\ref{sec-3} details the derivation of unpolarized scattering amplitudes and the global fit results compared against TOTEM, D\O, and Tevatron data. Concluding remarks and an outlook are provided in Sec.~\ref{sec-4}.
}

\section{Theoretical Framework}\label{sec-2}

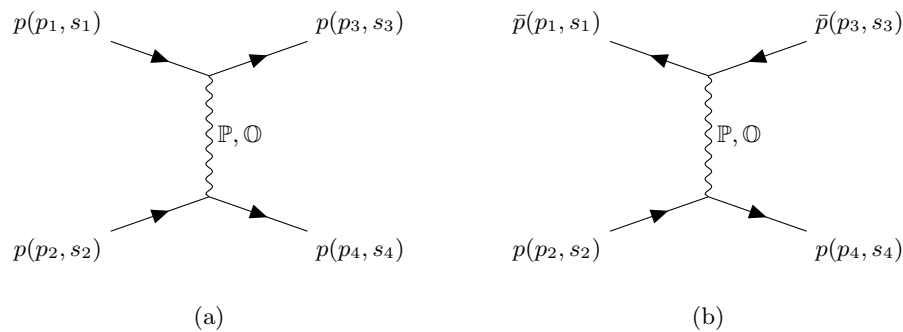
\begin{figure}[h]
\centering

\begin{tikzpicture}
\begin{feynman}
\vertex (p1) at (0,1.5) {\(p(p_1,s_1)\)};
\vertex (p2) at (0,-1.5) {\(p(p_2,s_2)\)};
\vertex (v1) at (2,0.8);
\vertex (v2) at (2,-0.8);
\vertex (p3) at (4,1.5) {\(p(p_3,s_3)\)};
\vertex (p4) at (4,-1.5) {\(p(p_4,s_4)\)};

\diagram*{
(p1) -- [fermion] (v1) -- [fermion] (p3),
(p2) -- [fermion] (v2) -- [fermion] (p4),
(v1) -- [boson, edge label={\(\mathbb{P},\mathbb{O}\)}] (v2),
};
\end{feynman}
\node at (2,-2.4) {(a)};
\end{tikzpicture}
\hspace{1cm}
\begin{tikzpicture}
\begin{feynman}
\vertex (pb1) at (0,1.5) {\(\bar p(p_1,s_1)\)};
\vertex (p2)  at (0,-1.5) {\(p(p_2,s_2)\)};
\vertex (v1)  at (2,0.8);
\vertex (v2)  at (2,-0.8);
\vertex (pb3) at (4,1.5) {\(\bar p(p_3,s_3)\)};
\vertex (p4)  at (4,-1.5) {\(p(p_4,s_4)\)};

\diagram*{
(v1) -- [fermion] (pb1),
(pb3) -- [fermion] (v1),
(p2) -- [fermion] (v2) -- [fermion] (p4),
(v1) -- [boson, edge label={\(\mathbb{P},\mathbb{O}\)}] (v2),
};
\end{feynman}
\node at (2,-2.4) {(b)};
\end{tikzpicture}

\caption{Elastic scattering via Pomeron and Odderon exchange: 
(a) \(pp \to pp\), 
(b) \(p\bar p \to p\bar p\).}
\label{fig:elastic_PO}
\end{figure}

\subsection{Effective Lagrangians of the $pp$ and $p\bar p$ scattering in spin-2 pomeron and spin-3 odderon exchange picture}

In this section, we present the effective model Lagrangians for elastic $pp$ and $p\bar{p}$
 scattering mediated by spin-2 Pomeron and spin-3 Odderon exchange, as shown in Fig.~1. The explicit forms of these Lagrangians are adopted from \cite{magallanes2024odderon} and read

\begin{eqnarray}
\mathscr{L}_{\mathbb{P}} &=& -ig_{\mathbb{P}NN}\mathbb{P}^{\mu\nu}\mathcal{G}^{\mu'\nu'}_{\mu\nu}\bar{\psi}\gamma_{\mu'}\overleftrightarrow{\partial}_{\nu'}\psi 
\end{eqnarray}

\begin{eqnarray}
\mathscr{L}_{\mathbb{O}} &=& \frac{i}{3\,!}\,\frac{g_{\mathbb{O}NN}}{\mathcal{M}_0}\,\mathbb{O}^{\mu\nu\rho}\,\overline{\psi}\gamma_{\mu'}\,\overleftrightarrow{\partial}_{\nu'}\,\overleftrightarrow{\partial}_{\rho'}\,\psi\,\mathcal{G}^{\mu'\nu'\rho'}_{\mu\nu\rho}\,. 
\end{eqnarray}

Here, $\mathbb{P}^{\mu\nu}$ and $\mathbb{O}^{\mu\nu\rho}$ denote the spin--2 and spin--3 tensor fields associated with the Pomeron and Odderon, respectively. The parameter $\mathcal{M}_0$ is a mass scale fixed to $\mathcal{M}_0 = 1~\mathrm{GeV}$. The field $\psi$ represents the Dirac proton field, and the derivative operator is defined as $\overleftrightarrow{\partial}_{\nu} = \overleftarrow{\partial}_{\nu} - \overrightarrow{\partial}_{\nu}$. The coupling constants $g_{\mathbb{P}NN}$ and $g_{\mathbb{O}NN}$ correspond to the Pomeron--proton and Odderon--proton interactions, respectively. \textcolor{black}{The Pomeron coupling $g_{\mathbb{P}NN}$ here is related to the dimensionless coupling $\beta_{\mathbb{P}NN}$ of Ewerz, Maniatis, and Nachtmann (EMN)~\cite{ewerz2014model} by $g_{\mathbb{P}NN} = 2\,\mathcal{M}_0\,\beta_{\mathbb{P}NN}$, where the factor of $2$ arises from the symmetric tensor normalization. Numerical values quoted in this work are consistent with the EMN convention after this rescaling.}
The totally symmetric tensor operators for the Pomeron and Odderon exchanges are defined as
\begin{equation}
\mathcal{G}^{\mu'\nu'}_{\mu\nu}
= \frac{1}{2!}
\left(
g^{\mu'}_{\mu} g^{\nu'}_{\nu}
+ g^{\mu'}_{\nu} g^{\nu'}_{\mu}
\right),
\end{equation}
and
\begin{equation}
\mathcal{G}^{\mu'\nu'\rho'}_{\mu\nu\rho}
= \frac{1}{3!}
\Big(
g^{\mu'}_{\mu} g^{\nu'}_{\nu} g^{\rho'}_{\rho}
+ g^{\mu'}_{\nu} g^{\nu'}_{\rho} g^{\rho'}_{\mu}
+ g^{\mu'}_{\rho} g^{\nu'}_{\mu} g^{\rho'}_{\nu}
+ g^{\mu'}_{\nu} g^{\nu'}_{\mu} g^{\rho'}_{\rho}
+ g^{\mu'}_{\mu} g^{\nu'}_{\rho} g^{\rho'}_{\nu}
+ g^{\mu'}_{\rho} g^{\nu'}_{\nu} g^{\rho'}_{\mu}
\Big).
\end{equation}
Throughout this work, we adopt the metric convention
\(g_{\mu\nu} = \mathrm{diag}(1,-1,-1,-1)\).

\subsection{Elastic Scattering Amplitudes }
After specifying the effective interaction Lagrangians, the elastic scattering amplitudes are constructed using standard quantum field theory techniques applied within an effective Regge framework\cite{peskin2018introduction,das2020lectures}.In the Regge limit, the resulting invariant amplitudes for $pp$ and $p\bar{p}$ elastic scattering, including both Pomeron and Odderon exchanges, can be expressed as

\begin{eqnarray}
i\mathcal{A}^{pp}_{\mathbb{P}}\,&&=\langle p(q_3)p(q_4)|:\mathcal{T}exp \Bigg(i\int \mathcal{L}_\mathbb{P}(x)d^4x\Bigg):|p(q_1)p(q_2)\rangle\nonumber\\
&&= -ig^2_{\mathbb{P}NN}\mathcal{G}^{\mu'_1\nu'_1}_{\mu_1\nu_1}\bar{u}(q_3)\gamma_{\mu'_1}(q_{3,\nu'_1}+q_{1,\nu'_1})u(q_1)i\Delta^{\mu_1\nu_1;\mu_2\nu_2}_{\mathbb{P}}(s,t)\,\mathcal{G}^{\mu'_2\nu'_2}_{\mu_2\nu_2}\bar{u}(q_4)(q_{4,\nu'_2}+q_{2,\nu'_2})u(q_2)F^2_{\mathbb{P}}(t),
\end{eqnarray}

\begin{eqnarray}
i\mathcal{A}^{p\bar{p}}_{\mathbb{P}}\,&&=\langle \bar{p}(q_3)p(q_4)|:\mathcal{T}exp \Bigg(i\int \mathcal{L}_\mathbb{P}(x)d^4x\Bigg):|\bar{p}(q_1)p(q_2)\rangle\nonumber\\
&&= -ig^2_{\mathbb{P}NN}\mathcal{G}^{\mu'_1\nu'_1}_{\mu_1\nu_1}\bar{v}(q_1)\gamma_{\mu'_1}(q_{1,\nu'_1}+q_{3,\nu'_1})v(q_3)i\Delta^{\mu_1\nu_1;\mu_2\nu_2}_{\mathbb{P}}(s,t)\,\mathcal{G}^{\mu'_2\nu'_2}_{\mu_2\nu_2}\bar{u}(q_4)(q_{4,\nu'_2}+q_{2,\nu'_2})u(q_2)F^2_{\mathbb{P}}(t),
\end{eqnarray}

\begin{eqnarray}
i\mathcal{A}^{pp}_{\mathbb{O}}\,&&=\langle p(q_3)p(q_4)|:\mathcal{T}exp \Bigg(i\int \mathcal{L}_\mathbb{O}(x)d^4x\Bigg):|p(q_1)p(q_2)\rangle\nonumber\\
&&=\,\frac{ig^{2}_{\mathbb{O}NN}}{36\mathcal{M}^2_0}\,\overline{u}\,(q_{3})\gamma_{\mu'_{1}}\,(q_{3,\nu'_1}+q_{1,\nu'_1})\,(q_{3,\rho'_1}+q_{1,\rho'_1})\,u(q_{1})\,\mathcal{G}^{\mu'_{1}\nu'_{1}\rho'_{1}}_{\mu_{1}\nu_{1}\rho_{1}}\,\Delta^{\mu_{1}\nu_{1}\rho_{1};\mu_{2}\nu_{2}\rho_{2}}_{\mathbb{O}}\,(s,t)\,\mathcal{G}^{\mu'_{2}\nu'_{2}\rho'_{2}}_{\mu_{2}\nu_{2}\rho_{2}}\nonumber\\
&&\quad\,\times\,\overline{u}\,(q_{4})\,\gamma_{\mu'_{2}}\,(q_{4,\nu'_2}+q_{2,\nu'_2})\,(q_{4,\rho'_2}+q_{2,\rho'_2})\,u(q_{2})\,\rev{F^{2}_{\mathbb{O}}\,(t).} 
\end{eqnarray}

\begin{eqnarray}
i\mathcal{A}^{p\bar{p}}_{\mathbb{O}}\,&&=\langle \bar{p}(q_3)p(q_4)|:\mathcal{T}exp \Bigg(i\int \mathcal{L}_\mathbb{O}(x)d^4x\Bigg):|\bar{p}(q_1)p(q_2)\rangle\nonumber\\
&&=\,\frac{ig^{2}_{\mathbb{O}NN}}{36\mathcal{M}^2_0}\,\rev{\overline{v}\,(q_{1})\gamma_{\mu'_{1}}\,(q_{1,\nu'_1}+q_{3,\nu'_1})\,(q_{1,\rho'_1}+q_{3,\rho'_1})\,v(q_{3})}
\,\mathcal{G}^{\mu'_{1}\nu'_{1}\rho'_{1}}_{\mu_{1}\nu_{1}\rho_{1}}\,\Delta^{\mu_{1}\nu_{1}\rho_{1};\mu_{2}\nu_{2}\rho_{2}}_{\mathbb{O}}\,(s,t)\,\mathcal{G}^{\mu'_{2}\nu'_{2}\rho'_{2}}_{\mu_{2}\nu_{2}\rho_{2}}\nonumber\\
&&\quad\,\times\,\overline{u}\,(q_{4})\,\gamma_{\mu'_{2}}\,(q_{4,\nu'_2}+q_{2,\nu'_2})\,(q_{4,\rho'_2}+q_{2,\rho'_2})\,u(q_{2})\,\rev{F^{2}_{\mathbb{O}}\,(t).} 
\end{eqnarray}

Here $i\Delta^{\mu_1\nu_1;\mu_2\nu_2}_{\mathbb{P}}(t,s)$ and
$i\Delta^{\mu_1\nu_1\rho_1;\mu_2\nu_2\rho_2}_{\mathbb{O}}(t,s)$ denote the
propagators of the tensor Pomeron and the spin--3 Odderon, respectively.
In the present work, these propagators are modeled as effective massive
exchanges and written in factorized form as a Lorentz projector multiplied
by a scalar propagator.

The tensor Pomeron propagator is given by
\begin{eqnarray}
i\Delta^{\mu_{1}\nu_{1};\mu_{2}\nu_{2}}_{\mathbb{P}}(t,s)
&=&
\mathcal{P}^{\mu_{1}\nu_{1};\mu_{2}\nu_{2}}\,
\tilde{\Delta}_{\mathbb{P}}(t,s),
\end{eqnarray}
with the scalar part
\begin{eqnarray}
\tilde{\Delta}_{\mathbb{P}}(t)
&=&
\frac{1}{t - M_{\mathbb{P}}^{2}},
\end{eqnarray}
and the spin--2 projector
\begin{eqnarray}
\mathcal{P}^{\mu_{1}\nu_{1};\mu_{2}\nu_{2}}
&=&
\frac{1}{2}
\left(
P^{\mu_{1}\mu_{2}} P^{\nu_{1}\nu_{2}}
+
P^{\mu_{1}\nu_{2}} P^{\nu_{1}\mu_{2}}
\right)
-
\frac{1}{3}
P^{\mu_{1}\nu_{1}} P^{\mu_{2}\nu_{2}} .
\end{eqnarray}

Analogously, the Odderon propagator is written as
\begin{eqnarray}
i\Delta^{\mu_{1}\nu_{1}\rho_{1};\mu_{2}\nu_{2}\rho_{2}}_{\mathbb{O}}(t,s)
&=&
\mathcal{P}^{\mu_{1}\nu_{1}\rho_{1};\mu_{2}\nu_{2}\rho_{2}}\,
\tilde{\Delta}_{\mathbb{O}}(t,s),
\end{eqnarray}
where the scalar propagator is
\begin{eqnarray}
\tilde{\Delta}_{\mathbb{O}}(t)
&=&
\frac{1}{t - M_{\mathbb{O}}^{2}} .
\end{eqnarray}

The rank--three spin--3 projector is given by
\begin{eqnarray}
\mathcal{P}^{\mu_1\nu_1\rho_1;\mu_2\nu_2\rho_2}
&=&
\frac{1}{6}
\Big[
P^{\mu_1\mu_2}P^{\nu_1\nu_2}P^{\rho_1\rho_2}
+
P^{\mu_1\mu_2}P^{\nu_1\rho_2}P^{\rho_1\nu_2}
+
P^{\mu_1\nu_2}P^{\nu_1\mu_2}P^{\rho_1\rho_2}
\nonumber\\
&&\quad
+
P^{\mu_1\nu_2}P^{\nu_1\rho_2}P^{\rho_1\mu_2}
+
P^{\mu_1\rho_2}P^{\nu_1\mu_2}P^{\rho_1\nu_2}
+
P^{\mu_1\rho_2}P^{\nu_1\nu_2}P^{\rho_1\mu_2}
\Big]
\nonumber\\
&-&
\frac{1}{15}
\Big[
P^{\mu_1\nu_1}P^{\mu_2\nu_2}P^{\rho_1\rho_2}
+
P^{\mu_1\rho_1}P^{\mu_2\nu_2}P^{\nu_1\rho_2}
+
P^{\nu_1\rho_1}P^{\mu_2\nu_2}P^{\mu_1\rho_2}
\nonumber\\
&&\quad
+
P^{\mu_1\nu_1}P^{\mu_2\rho_2}P^{\rho_1\nu_2}
+
P^{\mu_1\rho_1}P^{\mu_2\rho_2}P^{\nu_1\nu_2}
+
P^{\mu_1\rho_1}P^{\nu_2\rho_2}P^{\nu_1\mu_2}
\nonumber\\
&&\quad
+
P^{\nu_1\rho_1}P^{\nu_2\rho_2}P^{\mu_1\mu_2}
\Big] .
\end{eqnarray}

The tensor
\begin{eqnarray}
P^{\mu\nu}(k)
&=&
- g^{\mu\nu} + \frac{k^\mu k^\nu}{M_{\mathbb{O}}^{2}}
\end{eqnarray}
ensures transversality of the exchanged states.
Here $M_{\mathbb{P}}$ and $M_{\mathbb{O}}$ denote the effective masses of
the Pomeron and Odderon, respectively.

\subsection{Reggeization of Scattering Amplitudes}

{\color{black}
To preserve $s$-channel unitarity and crossing symmetry at high energies, we implement the Reggeization of $t$-channel exchange amplitudes following the prescription of Ref.~\cite{mathieu2015neutral}. A fixed-spin $J \geq 2$ exchange generates an amplitude growing as $s^{J-1}$, which violates the Froissart--Martin bound $\sigma_{\rm tot} \leq C \ln^2 s$ as $s \to \infty$. Reggeization replaces the fixed integer spin with a continuous trajectory $\alpha_X(t)$ whose intercept lies near unity, effectively restoring consistency with unitarity while preserving the $t$-channel structure of the underlying field-theoretic vertex. Analytically, Regge poles arise from the continuation of $t$-channel partial waves into the complex angular-momentum plane, simultaneously controlling the asymptotic high-energy behavior in the $s$-channel and the spectrum of resonances at $J = \alpha_X(m^2)$. This procedure dynamically generates the correct charge-conjugation structure via the signature factors 
\begin{equation}
    \eta_X(t) = \frac{1}{2}\left[(-1)^{J_X} + e^{-i\pi\alpha_X(t)}\right].
\end{equation}
The relative phase fixed by these factors determines the sign of the interference term between the even-signature Pomeron ($C=+1$) and the odd-signature Odderon ($C=-1$), enabling a \rev{Regge-theory prediction of the relative phase controlling the $pp$ and $p\bar{p}$ dip--bump asymmetry.} 

For a particle $X$ with mass $m_X$ and spin $J_X$, the Reggeized propagator $\mathcal{R}_X(s,t)$ takes the generic form
\begin{equation}
\mathcal{R}_X(s,t) = \frac{\pi \alpha'_X \eta_X(t)}{\Gamma[\alpha_X(t) + 1 - J_X] \sin[\pi \alpha_X(t)]} \left(\frac{s}{s_0}\right)^{\alpha_X(t) - J_X},
\end{equation}
where the Regge trajectory is linearized as $\alpha_X(t) = J_X + \alpha'_X(t - m_X^2)$ and $s_0 = 1~\text{GeV}^2$. Implementing this for the spin-2 Pomeron exchange ($\mathbb{P}$), the $t$-channel Feynman propagator is replaced by the Reggeized kernel
\begin{equation}
\tilde{\Delta}_{\mathbb{P}}(t) \xrightarrow{J=2} \mathcal{R}_{\mathbb{P}}(s,t) = \frac{\pi \alpha'_{\mathbb{P}} \cdot \frac{1}{2}[1 + e^{-i\pi\alpha_{\mathbb{P}}(t)}]}{\Gamma[\alpha_{\mathbb{P}}(t) - 1] \sin[\pi \alpha_{\mathbb{P}}(t)]} \left(\frac{s}{s_0}\right)^{\alpha_{\mathbb{P}}(t) - 2},
\end{equation}
with $\alpha_{\mathbb{P}}(t) = 2 + \alpha'_{\mathbb{P}}(t - M_{\mathbb{P}}^2)$, utilizing the numerical values $\alpha'_{\mathbb{P}} = 0.25~\text{GeV}^{-2}$ and $M_{\mathbb{P}} = 1.918~\text{GeV}$. Similarly, for the spin-3 Odderon exchange ($\mathbb{O}$), the replacement yields
\begin{equation}
\tilde{\Delta}_{\mathbb{O}}(t) \xrightarrow{J=3} \mathcal{R}_{\mathbb{O}}(s,t) = \frac{\pi \alpha'_{\mathbb{O}} \cdot \frac{1}{2}[-1 + e^{-i\pi\alpha_{\mathbb{O}}(t)}]}{\Gamma[\alpha_{\mathbb{O}}(t) - 2] \sin[\pi \alpha_{\mathbb{O}}(t)]} \left(\frac{s}{s_0}\right)^{\alpha_{\mathbb{O}}(t) - 3},
\end{equation}
where $\alpha_{\mathbb{O}}(t) = 3 + \alpha'_{\mathbb{O}}(t - M_{\mathbb{O}}^2)$, with the averaged inputs $\alpha'_{\mathbb{O}} = 0.189~\text{GeV}^{-2}$ and $M_{\mathbb{O}} = 3.201~\text{GeV}$ from Ref.~\cite{magallanes2024odderon}. 

The resulting Pomeron intercept $\alpha_{\mathbb{P}}(0) \approx 2 - \alpha'_{\mathbb{P}} M_{\mathbb{P}}^2 \approx 1.30$ lies above the canonical soft-Pomeron value ($\approx 1.08$) and below the LO BFKL hard-Pomeron value ($\approx 1.4$). \rev{In the present analysis this value should not be interpreted as a universal soft-Pomeron intercept extracted from total cross sections. It is an effective intercept associated with the restricted differential-cross-section fit and with the Born-level tensor-exchange ansatz.} Unlike previous models that incorporate Regge behavior directly into an effective propagator, our framework utilizes explicit spin projectors alongside Reggeization to ensure a transparent treatment of the tensor structure and form-factor dependence, which is critical for extracting the Odderon-vertex transverse profile in the following sections.}

\subsection{Hadronic Form Factors}

{\color{black}
In this section, we define the hadronic form factors used to parametrize the extended structure of the proton within the Regge limit, characterized by high center-of-mass energies and small momentum transfers ($|t| \ll s$). In this kinematic regime, these functions effectively model the non-local distribution of the proton's interaction with Reggeized exchanges, providing a phenomenological suppression of unphysical large-$|t|$ contributions without necessitating a full description of short-distance dynamics. For the Pomeron exchange, we adopt the standard proton--Pomeron coupling form factor widely used in tensor-Pomeron phenomenology \cite{ewerz2014model, magallanes2024odderon, sawasdipol2023effects}, defined as:
\begin{equation}
F_{\mathbb{P}}(t) = \left(1-\frac{t}{4m_p^2}\frac{\mu_p}{\mu_N}\right) \left(1-\frac{t}{4m_p^2}\right)^{-1} \left(1-\frac{t}{m_D^2}\right)^{-2},
\end{equation}
where $\mu_N = e/(2m_p)$ is the nuclear magneton, $\mu_p/\mu_N = 2.7928$ is the proton magnetic moment, and $m_D^2 = 0.71~\mathrm{GeV}^2$ is the dipole mass parameter.

\rev{In the physical elastic region we use $t=q^2<0$ and $|t|=-t$. For a meaningful comparison of different vertex shapes, the Odderon form factors are treated as phenomenological functions of the momentum transfer and are normalized consistently at the forward point, $F_{\mathbb O}(0)=1$, unless stated otherwise. This convention prevents the comparison among FF1--FF7 from being driven by trivial normalization offsets rather than by genuine differences in $t$ dependence.}

Regarding the Odderon exchange, we examine seven candidate form factors (FF1--FF7) motivated by the non-local vertex structures discussed in Ref.~\cite{mart2013hadronic}. These parametrizations are selected to investigate different momentum-transfer behaviors while maintaining a lower parameter count than previous multi-component studies. The first candidate, FF1, is a standard dipole form factor providing power-law suppression based on the off-shellness of the exchange:
\begin{equation}
F_{1}(q^{2}) = \frac{\Delta^{4}}{\Delta^{4} + \left(q^{2} - M_{\mathbb{O}}^{2}\right)^{2}}.
\end{equation}
To model a more compact interaction region with steeper suppression, we consider a quartic dipole form factor (FF2), characterized by a fixed fourth-order power-law tail:
\begin{equation}
F_{2}(q^{2}) = \left[ 1 + \left( \dfrac{q^{2} - M_{\mathbb{O}}^{2}}{\Delta^{2}} \right)^{4} \right]^{-1}.
\end{equation}
A more flexible extension is provided by the generalized polynomial series (FF3), which incorporates a finite sum to account for possible multi-scale vertex substructures:
\begin{equation}
F_{3}(q^{2}) = \left[ 1 + \sum_{\alpha=1}^{n} \left| \dfrac{q^{2} - M_{\mathbb{O}}^{2}}{\Delta^{2}} \right|^{\alpha} \right]^{-1}.
\end{equation}
Alternatively, a smooth, Gaussian-type suppression of the vertex coupling can be modeled using the Gaussian form factor (FF4):
\begin{equation}
F_{4}(q^{2}) = \exp\!\left[ -\frac{\left(q^{2} - M_{\mathbb{O}}^{2}\right)^{2}}{\Delta^{4}} \right].
\end{equation}
Hybrid structures are also examined, such as the combined Gaussian--generalized dipole form (FF5), which merges Gaussian damping with a power-law tail governed by a free parameter $n$:
\begin{equation}
F_{5}(q^{2}) = \frac{ \exp\!\left[ -\dfrac{\left(q^{2} - M_{\mathbb{O}}^{2}\right)^{2}}{\Delta^{4}} \right] }{ 1 + \left( \dfrac{q^{2} - M_{\mathbb{O}}^{2}}{\Delta^{2}} \right)^{n} }.
\end{equation}
Similarly, a combined Gaussian--polynomial parametrization (FF6) is utilized to allow for more nuanced behavior in the intermediate-$|t|$ region:
\begin{equation}
F_{6}(q^{2}) = \frac{ \exp\!\left[ -\dfrac{\left(q^{2} - M_{\mathbb{O}}^{2}\right)^{2}}{\Delta^{4}} \right] }{ 1 + \sum_{\alpha=1}^{n} \left| \dfrac{q^{2} - M_{\mathbb{O}}^{2}}{\Delta^{2}} \right|^{n} }.
\end{equation}
Finally, we consider a purely geometric exponential $t$-slope form factor (FF7) based on the Mandelstam variable $t \equiv q^2$, assuming a single transverse interaction scale:
\begin{equation}
F_{7}(t) = \exp\!\left( -\frac{1}{2}B|t| \right).
\end{equation}

In the expressions above, $q$ denotes the four-momentum transfer and $M_{\mathbb{O}}$ represents an \rev{effective mass scale entering the Reggeized Odderon kernel rather than a direct determination of a physical glueball mass}. The cutoff scale $\Delta$ is associated with vertex compositeness and is fixed at $1~\mathrm{GeV}$ to ensure model parsimony. The parameter $B$ denotes the exponential slope, while $n$ dictates the order or power in the generalized forms. Note that for FF3 and FF6, the summation index $\alpha$ spans the polynomial terms up to order $n$.}

\subsection{Differential Cross-Section}

{\color{black}
The primary observable analyzed in this work is the unpolarized differential cross section for elastic $pp$ and $p\bar{p}$ scattering in the high-energy Regge limit. In accordance with standard relativistic kinematics, the differential cross section is related to the invariant scattering amplitude $\mathcal{A}(s,t)$ by
\begin{equation}
\frac{d\sigma^{pp, p\bar{p}}}{dt} = \frac{1}{64\pi s (s - 4m_p^2)} \sum_{\mathrm{spins}} \left|\mathcal{A}^{pp, p\bar{p}}(s,t)\right|^2,
\end{equation}
where the prefactor includes the $1/4$ spin-averaging constant for the initial-state protons and $m_p$ denotes the proton mass. The total scattering amplitude is constructed as the coherent sum of the Pomeron ($\mathbb{P}$) and Odderon ($\mathbb{O}$) contributions,
\begin{equation}
\mathcal{A}^{pp/p\bar{p}}(s,t) = \mathcal{A}_{\mathbb{P}}(s,t) \mp \mathcal{A}_{\mathbb{O}}(s,t),
\end{equation}
where the upper (lower) sign corresponds to $pp$ ($p\bar{p}$) scattering. This sign convention explicitly reflects the opposite charge-conjugation properties of the exchanges ($C_{\mathbb{P}}=+1$ and $C_{\mathbb{O}}=-1$), which dictate the relative phase between the $pp$ and $p\bar{p}$ channels.

The spin-summed squared amplitude, accounting for interference between the two $C$-parity sectors, is given by
\begin{align}
\sum_{\mathrm{spins}} \left|\mathcal{A}^{pp/p\bar{p}}\right|^2 = \sum_{\mathrm{spins}} \Big( \left|\mathcal{A}_{\mathbb{P}}\right|^2 + \left|\mathcal{A}_{\mathbb{O}}\right|^2 \mp 2\,\mathrm{Re} \left[ \mathcal{A}_{\mathbb{P}}^{*} \mathcal{A}_{\mathbb{O}} \right] \Big).
\end{align}

A crucial feature of this framework is that the relative phase between $\mathcal{A}_{\mathbb{P}}$ and $\mathcal{A}_{\mathbb{O}}$ is not an unconstrained phenomenological parameter. Instead, it is strictly determined by the Reggeized signature factors within the respective propagators. These factors fix the analytic structure and the ratio of real to imaginary parts for each amplitude as a function of the momentum transfer $t$. Consequently, the interference term which is responsible for the characteristic dip--bump differences between $pp$ and $p\bar{p}$ data is \rev{fixed within the chosen Regge-pole convention once the trajectory parameters and $C$-parity assignments are specified.}}

\section{Results and Discussions}\label{sec-3}
\subsection{Analytical Evaluation and Spin Sums}
Having established the theoretical framework, we evaluate the differential cross-sections by performing the large spin sums for the invariant amplitudes. Due to the significant algebraic complexity introduced by the rank-3 tensor Odderon and rank-2 Pomeron projectors, symbolic computations were performed using Mathematica \cite{Wolfram2019} and the FeynCalc package \cite{Shtabovenko2020,Shtabovenko2016,Mertig1991}. We impose the Regge limit by retaining only the leading-order contributions in the center-of-mass energy $\sqrt{s}$.The resulting analytical expressions for the $pp$ and $p\bar{p}$ channels exhibit the characteristic power-law scaling in $s$ and the oscillatory behavior dictated by the Regge signature factors and Gamma functions. Notably, the relative phases and signs between the $C=+1$ Pomeron and $C=-1$ Odderon are strictly preserved in this coherent summation, ensuring the physical integrity of the interference term which governs the dip-bump structure. Under these assumptions, the resulting expressions read

\begin{eqnarray}
\sum_{\rm spin}\,\big|\mathcal{A}_{\mathbb{P}}^{pp}\big|^2 &=& \sum_{\rm spin}\,\mathcal{A}_{\mathbb{P}}^{pp}\,\mathcal{A}_{\mathbb{P}}^{pp\,*} 
\nonumber\\
&=&\,\text{Tr}\Big[(\slashed{q_3}+m_p)\gamma_{\mu'_1}(q_{3,\nu'_1}+q_{1,\nu'_1})(\slashed{q_1}+m_p)\gamma_{\mu'_3}(q_{3,\nu'_3}+q_{1,\nu'_3})\nonumber\\
&\;&\quad\times\,(\slashed{q_4}+m_p)\gamma_{\mu'_2}(q_{4,\nu'_2}+q_{2,\nu'_2})(\slashed{q_2}+m_p)\gamma_{\mu'_4}(q_{4,\nu'_4}+q_{2,\nu'_4})\Big]\nonumber\\
&\;&\quad\times\,g^4_{\mathbb{P}}\mathcal{G}^{\mu'_1\nu'_1}_{\mu_1\nu_1}\mathcal{G}^{\mu'_2\nu'_2}_{\mu_2\nu_2}\mathcal{G}^{\mu'_3\nu'_3}_{\mu_3\nu_3}\mathcal{G}^{\mu'_4\nu'_4}_{\mu_4\nu_4}\Delta^{\mu_1\nu_1;\mu_2\nu_2}_{\mathbb{P}}(s,t)\Delta^{\mu_3\nu_3;\mu_4\nu_4}_{\mathbb{P}}(s,t)F^{4}_{\mathbb{P}}(t)\nonumber\\
&\;&\approx \frac{
631.655\, g_{\mathbb{P}NN}^{4}\, s^{4}
\left( \frac{s}{s_{0}} \right)^{ 2(-M_{\mathbb{P}}^{2}+t)\alpha_{\mathbb{P}}' }
\alpha_{\mathbb{P}}'^{2}\,F^{4}_{\mathbb{P}}(t)
}{
\Gamma^{2}\!\left(1 - M_{\mathbb{P}}^{2}\alpha_{\mathbb{P}}' + t\alpha_{\mathbb{P}}'\right)
}
\nonumber\\[0.3em]
& &
\times\;
\cos^{2}\!\left[1.5708\!\left(2 - M_{\mathbb{P}}^{2}\alpha_{\mathbb{P}}' + t\alpha_{\mathbb{P}}'\right)\right]\csc^{2}\!\left[
\pi\left(2 - M_{\mathbb{P}}^{2}\alpha_{\mathbb{P}}' + t\alpha_{\mathbb{P}}'\right)
\right]
\end{eqnarray}
\begin{eqnarray}
\sum_{\rm spin}\,\big|\mathcal{A}_{\mathbb{O}}^{pp}\big|^2 &=& \,\sum_{\rm spin}\,\mathcal{A}_{\mathbb{O}}^{pp}\,\mathcal{A}_{\mathbb{O}}^{pp\,*}
\nonumber\\
&=&\,\frac{g^{4}_{\mathbb{O}NN}}{1296}\,\text{Tr}\Big[(\slashed{q_{3}}+m_{p})\gamma_{\mu'_{1}}(q_{3,\nu'_{1}}+q_{1,\nu'_{1}})(q_{3,\rho'_{1}}+q_{1,\rho'_{1}})(\slashed{q_{1}}+m_{p})\gamma_{\bar{\mu'_{3}}}\nonumber\\
&\;&\quad\times\,(q_{3,\bar{\nu'_{3}}}+q_{1,\bar{\nu'_{3}}})(q_{3,\bar{\rho'_{3}}}+q_{1,\bar{\rho'_{3}}})(\slashed{q_{4}}+m_{p})\gamma_{\mu'_{2}}(q_{4,\nu'_{2}}+q_{2,\nu'_{2}})\nonumber\\
&\;&\quad\times\,(q_{4,\rho'_{2}}+q_{2,\rho'_{2}})(\slashed{q_{2}}+m_{p})\gamma_{\bar{\mu'_{4}}}(q_{4,\bar{\nu'_{4}}}+q_{2,\bar{\nu'_{4}}})(q_{4,\bar{\rho'_{4}}}+q_{2,\bar{\rho'_{4}}})\Big]\nonumber\\
&\;&\quad\times\,\mathcal{G}^{\mu'_{1}\nu'_{1}\rho'_{1}}_{\mu_{1}\nu_{1}\rho_{1}}\,\mathcal{G}^{\bar{\mu'_{3}}\bar{\nu'_{3}}\bar{\rho'_{3}}}_{\bar{\mu_{3}}\bar{\nu_{3}}\bar{\rho_{3}}}\,\mathcal{G}^{\mu'_{2}\nu'_{2}\rho'_{2}}_{\mu_{2}\nu_{2}\rho_{2}}\,\mathcal{G}^{\bar{\mu'_{4}}\bar{\nu'_{4}}\bar{\rho'_{4}}}_{\bar{\mu_{4}}\bar{\nu_{4}}\bar{\rho_{4}}}\,\Delta^{\mu_{1}\nu_{1}\rho_{1};\mu_{2}\nu_{2}\rho_{2}}_{\mathbb{O}}(s,t)\nonumber\\
&\;&\quad\times\,\Delta^{\bar{\mu_{3}}\bar{\nu_{3}}\bar{\rho_{3}};\bar{\mu_{4}}\bar{\nu_{4}}\bar{\rho_{4}}}_{\mathbb{O}}(s,t)F^{4}_{\mathbb{O}}(t)\nonumber\\
&\;&\approx \frac{
1.94955\, g_{\mathbb{O}NN}^{4}\, s^{6}
\left( \frac{s}{s_{0}} \right)^{ 2(-M_{\mathbb{O}}^{2}+t)\alpha_{\mathbb{O}}' }
\alpha_{\mathbb{O}}'^{2}\,F^{4}_{\mathbb{O}}(t)
}{
M^{4}_0\,
\Gamma^{2}\!\left(1 - M_{\mathbb{O}}^{2}\alpha_{\mathbb{O}}' + t\alpha_{\mathbb{O}}'\right)
}
\nonumber\\[0.3em]
& &
\times\;
\sin^{2}\!\left[
1.5708\left(3 - M_{\mathbb{O}}^{2}\alpha_{\mathbb{O}}' + t\alpha_{\mathbb{O}}'\right)
\right]\csc^{2}\!\left[\pi\left(M_{\mathbb{O}}^{2}-t\right)\alpha_{\mathbb{O}}'\right]
\end{eqnarray}

\begin{eqnarray}
\sum_{\rm spin}\,\big|\mathcal{A}_{\mathbb{P}}^{pp\,*}\,\mathcal{A}_{\mathbb{O}}^{pp}\big| 
&=& -\frac{g^{2}_{\mathbb{P}NN}\,g^{2}_{\mathbb{O}NN}}{164}\,\text{Tr}\Bigl\{\big[\gamma_{\bar{\mu_{3}}}(q_{3,\bar{\nu_{3}}}+q_{1,\bar{\nu_{3}}})+\gamma_{\bar{\nu_{3}}}(q_{3,\bar{\mu_{3}}}+q_{1,\bar{\mu_{3}}})\big](\slashed{q_{3}}+m_{p})\nonumber\\
&\;&\quad\times\,\gamma_{\mu'_{1}}(q_{3,\nu'_{1}}+q_{1,\nu'_{1}})(q_{3,\rho'_{1}}+q_{1,\rho'_{1}})(\slashed{q_{1}}+m_{p})\Bigl\}\nonumber\\
&\;&\quad\times\,\text{Tr}\Bigl\{\big[\gamma_{\bar{\mu_{4}}}(q_{4,\bar{\nu_{4}}}+q_{2,\bar{\nu_{4}}})+\gamma_{\bar{\nu_{4}}}(q_{4,\bar{\mu_{4}}}+q_{2,\bar{\mu_{4}}})\big](\slashed{q_{4}}+m_{p})\nonumber\\
&\;&\quad\times\,\gamma_{\mu'_{2}}(q_{4,\nu'_{2}}+q_{2,\nu'_{2}})(q_{4,\rho'_{2}}+q_{2,\rho'_{2}})(\slashed{q_{2}}+m_{p})\Bigl\}\nonumber\\
&\;&\quad\times\,\mathcal{G}^{\mu'_{1}\nu'_{1}\rho'_{1}}_{\mu_{1}\nu_{1}\rho_{1}}\,\mathcal{G}^{\mu'_{2}\nu'_{2}\rho'_{2}}_{\mu_{2}\nu_{2}\rho_{2}}\,\Delta^{\bar{\mu_{3}}\bar{\nu_{3}};\bar{\mu_{4}}\bar{\nu_{4}}}_{\mathbb{P}}(s,t)\,\Delta^{\mu_{1}\nu_{1}\rho_{1};\mu_{2}\nu_{2}\rho_{2}}_{\mathbb{O}}(s,t)F^{2}_{\mathbb{P}}(t)F^{2}_{\mathbb{O}}(t)\nonumber\\
&\;&\approx\;
\frac{32}{9}\,\pi^{2}\;
\Bigg|\,
\frac{
g_{\mathbb{O}NN}^{2}\, g_{\mathbb{P}NN}^{2}\, s^{5}
\left( \frac{s}{s_{0}} \right)^{
    -M_{\mathbb{O}}^{2}\alpha_{\mathbb{O}}' - M_{\mathbb{P}}^{2}\alpha_{\mathbb{P}}' + 
    t(\alpha_{\mathbb{O}}' + \alpha_{\mathbb{P}}')
}
\alpha_{\mathbb{O}}'\alpha_{\mathbb{P}}'\,F^{2}_{\mathbb{P}}(t)\,F^{2}_{\mathbb{O}}(t)
}{
M^{4}_0\,
\Gamma\!\left(1 - M_{\mathbb{O}}^{2}\alpha_{\mathbb{O}}' + t\alpha_{\mathbb{O}}'\right)\,
\Gamma\!\left(1 - M_{P}^{2}\alpha_{\mathbb{P}}' + t\alpha_{\mathbb{P}}'\right)
}
\nonumber\\[0.6em]
& &\qquad\qquad\times\;
\cos\!\left[
1.5708\left(2 - M_{\mathbb{P}}^{2}\alpha_{\mathbb{P}}' + t\alpha_{\mathbb{P}}'\right)
\right]\,\sin\!\left[
1.5708\left(3 - M_{\mathbb{O}}^{2}\alpha_{\mathbb{O}}' + t\alpha_{\mathbb{O}}'\right)
\right]
\nonumber\\[0.6em]
& &\qquad\qquad\times\;
\csc\!\left[
\pi\left(M_{\mathbb{O}}^{2} - t\right)\alpha_{\mathbb{O}}'
\right]\,
\csc\!\left[
\pi\left(2 - M_{\mathbb{P}}^{2}\alpha_{\mathbb{P}}' + t\alpha_{\mathbb{P}}'\right)
\right]\Bigg|
\end{eqnarray}
\begin{eqnarray}
\sum_{\rm spin}\,\big|\mathcal{A}_{\mathbb{P}}^{p\bar{p}}\big|^2 &=& \sum_{\rm spin}\,\mathcal{A}_{\mathbb{P}}^{p\bar{p}}\,\mathcal{A}_{\mathbb{P}}^{p\bar{p}\,*} 
\nonumber\\
&=&\,\text{Tr}\Big[(\slashed{q_1}-m_p)\gamma_{\mu'_1}(q_{3,\nu'_1}+q_{1,\nu'_1})(\slashed{q_3}-m_p)\gamma_{\mu'_3}(q_{3,\nu'_3}+q_{1,\nu'_3})\nonumber\\
&\;&\quad\times\,(\slashed{q_4}+m_p)\gamma_{\mu'_2}(q_{4,\nu'_2}+q_{2,\nu'_2})(\slashed{q_2}+m_p)\gamma_{\mu'_4}(q_{4,\nu'_4}+q_{2,\nu'_4})\Big]\nonumber\\
&\;&\quad\times\,g^4_{\mathbb{P}NN}\mathcal{G}^{\mu'_1\nu'_1}_{\mu_1\nu_1}\mathcal{G}^{\mu'_2\nu'_2}_{\mu_2\nu_2}\mathcal{G}^{\mu'_3\nu'_3}_{\mu_3\nu_3}\mathcal{G}^{\mu'_4\nu'_4}_{\mu_4\nu_4}\Delta^{\mu_1\nu_1;\mu_2\nu_2}_{\mathbb{P}}(s,t)\Delta^{\mu_3\nu_3;\mu_4\nu_4}_{\mathbb{P}}(s,t)F^4_{\mathbb{P}}(t)\nonumber\\
&\;&\approx\;\frac{
631.655\, g_{\mathbb{P}NN}^{4}\, s^{4}
\left( \frac{s}{s_{0}} \right)^{2(-M_{\mathbb{P}}^{2} + t)\alpha_{\mathbb{P}}'}
\alpha_{\mathbb{P}}'^{\,2}\,F^{4}_{\mathbb{P}}(t)
}{
\Gamma^{2}\!\left(1 - M_{\mathbb{P}}^{2}\alpha_{\mathbb{P}}' + t\alpha_{\mathbb{P}}'\right)
}
\nonumber\\[0.7em]
& &\quad\times\;
\cos^{2}\!\left[
1.5708\left(2 - M_{\mathbb{P}}^{2}\alpha_{\mathbb{P}}' + t\alpha_{\mathbb{P}}'\right)
\right]\csc^{2}\!\left[
\pi\left(2 - M_{\mathbb{P}}^{2}\alpha_{\mathbb{P}}' + t\alpha_{\mathbb{P}}'\right)
\right]
\end{eqnarray}

\begin{eqnarray}
\sum_{\rm spin}\,\big|\mathcal{A}_{\mathbb{O}}^{p\bar{p}}\big|^2 &=& \,\sum_{\rm spin}\,\mathcal{A}_{\mathbb{O}}^{p\bar{p}}\,\mathcal{A}_{\mathbb{O}}^{p\bar{p}\,*}
\nonumber\\
&=&\,\frac{g^{4}_{\mathbb{O}NN}}{1296}\,\text{Tr}\Big[(\slashed{q_{1}}-m_{p})\gamma_{\mu'_{1}}(q_{3,\nu'_{1}}+q_{1,\nu'_{1}})(q_{3,\rho'_{1}}+q_{1,\rho'_{1}})(\slashed{q_{3}}-m_{p})\gamma_{\bar{\mu'_{3}}}\nonumber\\
&\;&\quad\times\,(q_{3,\bar{\nu'_{3}}}+q_{1,\bar{\nu'_{3}}})(q_{3,\bar{\rho'_{3}}}+q_{1,\bar{\rho'_{3}}})(\slashed{q_{4}}+m_{p})\gamma_{\mu'_{2}}(q_{4,\nu'_{2}}+q_{2,\nu'_{2}})\nonumber\\
&\;&\quad\times\,(q_{4,\rho'_{2}}+q_{2,\rho'_{2}})(\slashed{q_{2}}+m_{p})\gamma_{\bar{\mu'_{4}}}(q_{4,\bar{\nu'_{4}}}+q_{2,\bar{\nu'_{4}}})(q_{4,\bar{\rho'_{4}}}+q_{2,\bar{\rho'_{4}}})\Big]\nonumber\\
&\;&\quad\times\,\mathcal{G}^{\mu'_{1}\nu'_{1}\rho'_{1}}_{\mu_{1}\nu_{1}\rho_{1}}\,\mathcal{G}^{\bar{\mu'_{3}}\bar{\nu'_{3}}\bar{\rho'_{3}}}_{\bar{\mu_{3}}\bar{\nu_{3}}\bar{\rho_{3}}}\,\mathcal{G}^{\mu'_{2}\nu'_{2}\rho'_{2}}_{\mu_{2}\nu_{2}\rho_{2}}\,\mathcal{G}^{\bar{\mu'_{4}}\bar{\nu'_{4}}\bar{\rho'_{4}}}_{\bar{\mu_{4}}\bar{\nu_{4}}\bar{\rho_{4}}}\,\Delta^{\mu_{1}\nu_{1}\rho_{1};\mu_{2}\nu_{2}\rho_{2}}_{\mathbb{O}}(s,t)\nonumber\\
&\;&\quad\times\,\Delta^{\bar{\mu_{3}}\bar{\nu_{3}}\bar{\rho_{3}};\bar{\mu_{4}}\bar{\nu_{4}}\bar{\rho_{4}}}_{\mathbb{O}}(s,t)F^{4}_{\mathbb{O}}(t)\nonumber\\
&\;&\approx\;
\frac{
1.94955\, g_{\mathbb{O}NN}^{4}\, s^{6}
\left( \frac{s}{s_{0}} \right)^{ 2(-M_{\mathbb{O}}^{2} + t)\alpha_{\mathbb{O}}' }
\alpha_{\mathbb{O}}'^{\,2}\,F^{4}_{\mathbb{O}}(t)
}{
M^{4}_0\,
\Gamma^{2}\!\left( 1 - M_{\mathbb{O}}^{2}\alpha_{\mathbb{O}}' + t\alpha_{\mathbb{O}}' \right)
}
\nonumber\\[0.5em]
& & \times 
\sin^{2}\!\left[ 1.5708\,(3 - M_{\mathbb{O}}^{2}\alpha_{\mathbb{O}}' + t\alpha_{\mathbb{O}}') \right]\csc^{2}\!\left[ \pi (M_{\mathbb{O}}^{2} - t)\alpha_{\mathbb{O}}' \right]
\end{eqnarray}

\begin{eqnarray}
\sum_{\rm spin}\,\big|\mathcal{A}_{\mathbb{P}}^{p\bar  {p}\,*}\,\mathcal{A}_{\mathbb{O}}^{p\bar{p}}\big| 
&=& -\frac{g^{2}_{\mathbb{P}NN}\,g^{2}_{\mathbb{O}NN}}{164}\,\text{Tr}\Bigl\{\big[\gamma_{\bar{\mu_{3}}}(q_{3,\bar{\nu_{3}}}+q_{1,\bar{\nu_{3}}})+\gamma_{\bar{\nu_{3}}}(q_{3,\bar{\mu_{3}}}+q_{1,\bar{\mu_{3}}})\big](\slashed{q_{3}}+m_{p})\nonumber\\
&\;&\quad\times\,\gamma_{\mu'_{1}}(q_{3,\nu'_{1}}+q_{1,\nu'_{1}})(q_{3,\rho'_{1}}+q_{1,\rho'_{1}})(\slashed{q_{1}}+m_{p})\Bigl\}\nonumber\\
&\;&\quad\times\,\text{Tr}\Bigl\{\big[\gamma_{\bar{\mu_{4}}}(q_{4,\bar{\nu_{4}}}+q_{2,\bar{\nu_{4}}})+\gamma_{\bar{\nu_{4}}}(q_{4,\bar{\mu_{4}}}+q_{2,\bar{\mu_{4}}})\big](\slashed{q_{4}}+m_{p})\nonumber\\
&\;&\quad\times\,\gamma_{\mu'_{2}}(q_{4,\nu'_{2}}+q_{2,\nu'_{2}})(q_{4,\rho'_{2}}+q_{2,\rho'_{2}})(\slashed{q_{2}}+m_{p})\Bigl\}\nonumber\\
&\;&\quad\times\,\mathcal{G}^{\mu'_{1}\nu'_{1}\rho'_{1}}_{\mu_{1}\nu_{1}\rho_{1}}\,\mathcal{G}^{\mu'_{2}\nu'_{2}\rho'_{2}}_{\mu_{2}\nu_{2}\rho_{2}}\,\Delta^{\bar{\mu_{3}}\bar{\nu_{3}};\bar{\mu_{4}}\bar{\nu_{4}}}_{\mathbb{P}}(s,t)\,\Delta^{\mu_{1}\nu_{1}\rho_{1};\mu_{2}\nu_{2}\rho_{2}}_{\mathbb{O}}(s,t)F^{2}_{\mathbb{P}}(t)F^{2}_{\mathbb{O}}(t)\nonumber\\
&\;&\approx\;\frac{32}{9}\,\pi^{2}\,
\Bigg|\;
\frac{
g_{\mathbb{O}}^{2}\, g_{\mathbb{P}NN}^{2}\, s^{5}
\left( \frac{s}{s_{0}} \right)^{ -M_{\mathbb{O}}^{2}\alpha_{\mathbb{O}}' 
    - M_{\mathbb{P}}^{2}\alpha_{\mathbb{P}}' 
    + t(\alpha_{\mathbb{O}}' + \alpha_{\mathbb{P}}')}
\alpha_{\mathbb{O}}'\, \alpha_{\mathbb{P}}'\,F^{2}_{\mathbb{P}}(t)\,F^{2}_{\mathbb{O}}(t)
}{
M^{4}_0\,
\Gamma\!\left( 1 - M_{\mathbb{O}}^{2}\alpha_{\mathbb{O}}' + t\alpha_{\mathbb{O}}' \right)\,
\Gamma\!\left( 1 - M_{\mathbb{P}}^{2}\alpha_{\mathbb{P}}' + t\alpha_{\mathbb{P}}' \right)
}
\nonumber\\[0.7em]
& &\qquad\qquad\times\;
\cos\!\left[ 1.5708\left(2 - M_{\mathbb{P}}^{2}\alpha_{\mathbb{P}}' + t\alpha_{\mathbb{P}}'\right) \right]\sin\!\left[ 1.5708 (3 - M_{\mathbb{O}}^{2}\alpha_{\mathbb{O}}' + t\alpha_{\mathbb{O}}') \right]
\nonumber\\[0.7em]
& &\qquad\qquad\times\;
\csc\!\left[\pi ( M_{\mathbb{O}
}^{2} - t )\alpha_{\mathbb{O}}'\right]\,
\csc\!\left[\pi ( 2 - M_{\mathbb{P}}^{2}\alpha_{\mathbb{P}}' + t\alpha_{\mathbb{P}}' )\right]
\Bigg|
\end{eqnarray}

\subsection{Global Fit Analysis and Form Factor Hierarchy}

\begin{figure}[h]
    \centering
    \includegraphics[width=\linewidth]{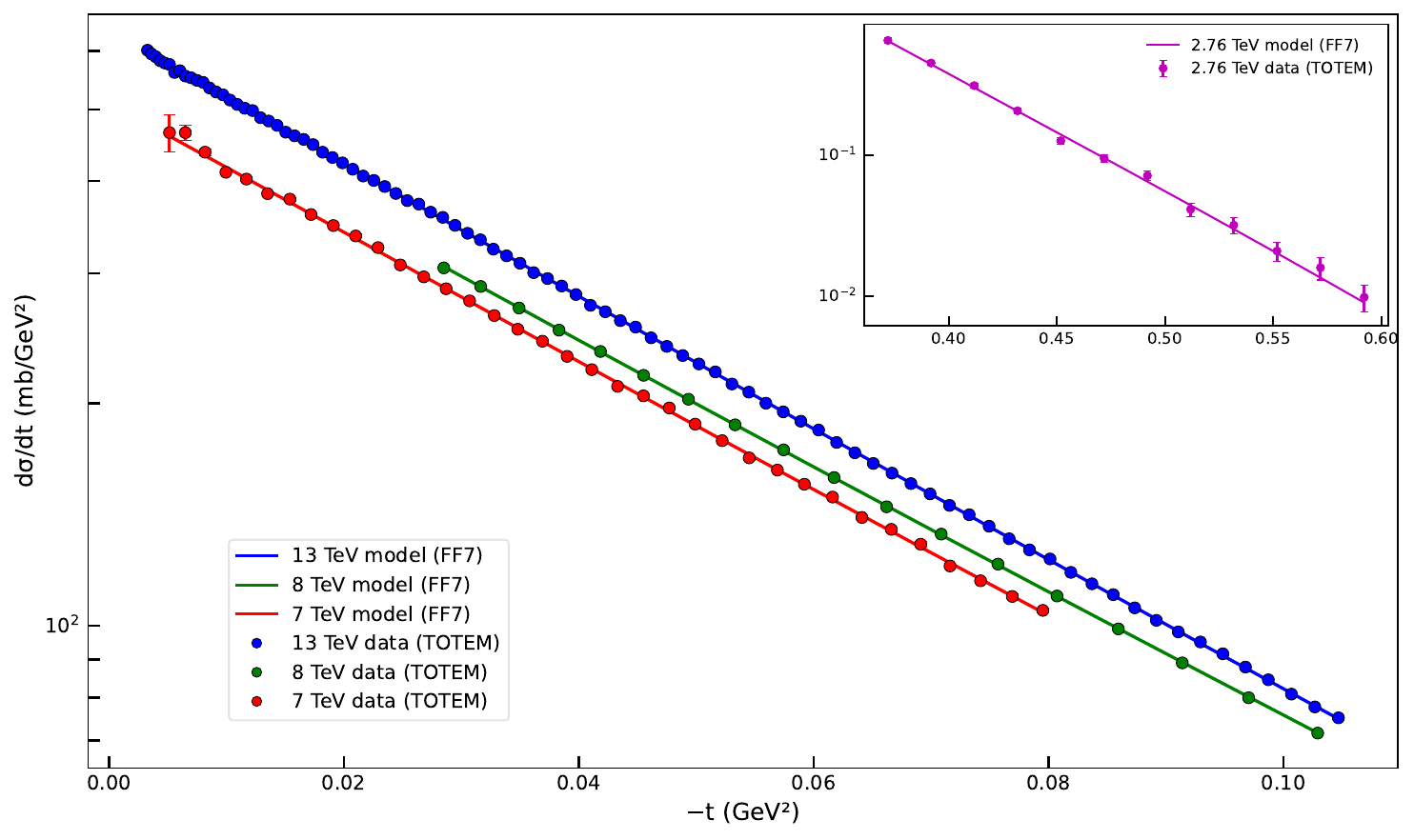}
    \caption{
     The differential cross section of elastic $pp$ scattering at
$\sqrt{s}=13$, 8, and 7~TeV compared with the global fit obtained using
the FF7 form factor.
The inset shows the corresponding result at $\sqrt{s}=2.76$~TeV.
    }
    \label{fig:globalfit_ff7}
\end{figure}

\begin{figure*}[h]
    \centering
    \includegraphics[width=0.75\textwidth]{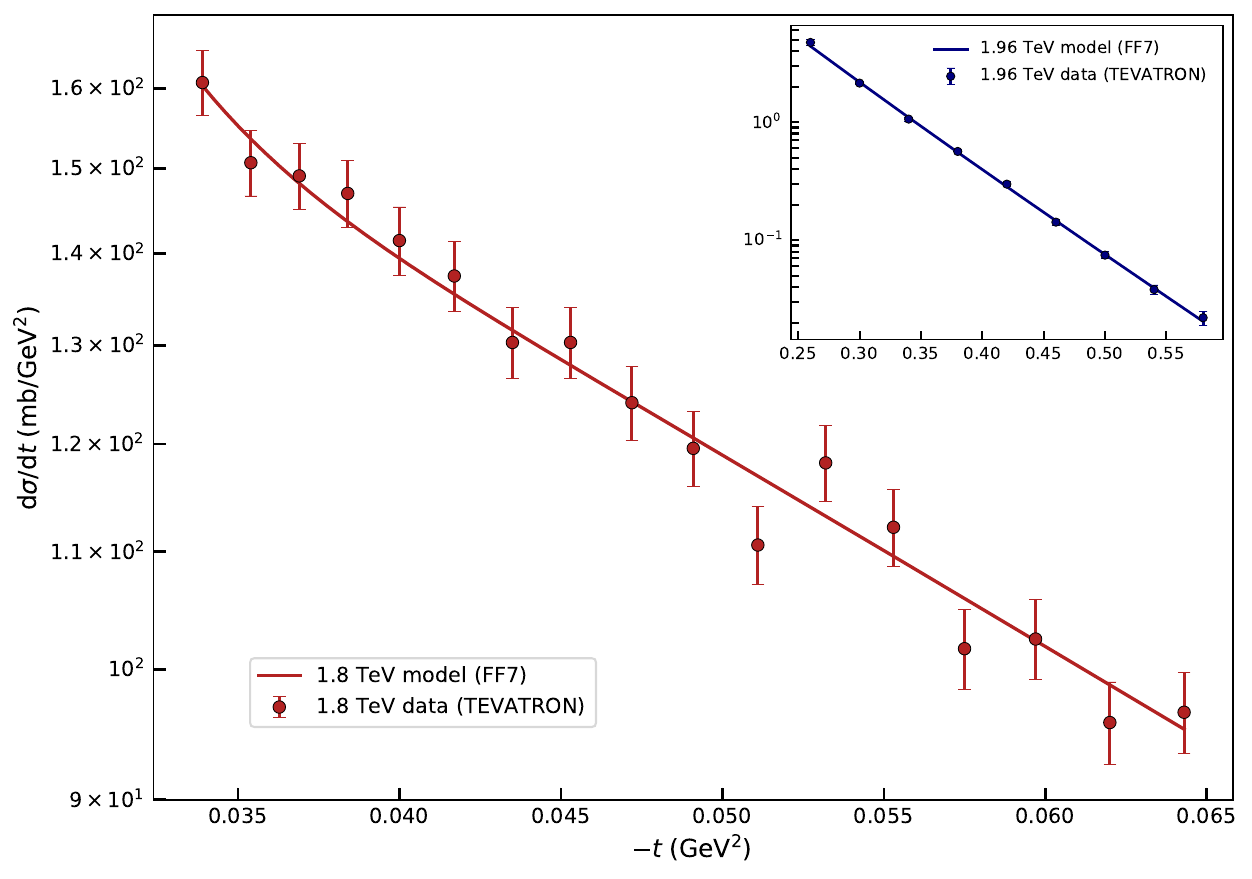}
    \caption{%
   The differential cross section for elastic $p\bar p$ scattering at
$\sqrt{s}=1.80~\mathrm{TeV}$ together with the global fit obtained using
the form factor FF7.
    The inset displays the corresponding comparison at
    $\sqrt{s}=1.96~\mathrm{TeV}$ using Tevatron data.
    }
    \label{fig:ppbar_ff7}
\end{figure*}

\begin{table}[!t]
\centering
\renewcommand{\arraystretch}{1.25}
\setlength{\tabcolsep}{8pt}

\caption{
Best-fit parameters of the FF7 model for elastic $pp$ scattering obtained from
Monte Carlo error propagation.
Uncertainties are statistical only.
The fit quality corresponds to a reduced $\chi^{2} \simeq 0.98$.
}

\label{tab:ff7_pp}

\begin{tabular}{lc}
\hline\hline
\textbf{Parameter} & \textbf{FF7} \\
\hline

$N_{13}$          & $0.0299 \pm 0.0076$ \\
$N_{8}$           & $0.0103 \pm 0.0073$ \\
$N_{7}$           & $0.0217 \pm 0.0098$ \\
$N_{2.76}$        & $16.0 \pm 9.3$ \\[2pt]

$\alpha'_{\mathbb{P}}(13~\mathrm{TeV})$ 
                  & $0.1913 \pm 0.0012$ \\
$\alpha'_{\mathbb{P}}(8~\mathrm{TeV})$  
                  & $0.1817 \pm 0.0044$ \\
$\alpha'_{\mathbb{P}}(7~\mathrm{TeV})$  
                  & $0.1859 \pm 0.0028$ \\
$\alpha'_{\mathbb{P}}(2.76~\mathrm{TeV})$ 
                  & $0.2183 \pm 0.0026$ \\[2pt]
$B_{13}~(\mathrm{GeV}^{-2})$
                  & $57.2194 \pm 19.1320$ \\

$B_{8}~(\mathrm{GeV}^{-2})$
                  & $40.6769 \pm 0.0021$ \\

$B_{7}~(\mathrm{GeV}^{-2})$
                  & $99.9242 \pm 0.0825$ \\

$B_{2.76}~(\mathrm{GeV}^{-2})$
                  & $36.0028 \pm 0.0003$ \\[2pt]                  

$g_{\mathbb{P}NN}$ & $5.61 \pm 0.21$ \\
$g_{\mathbb{O}NN}$  & $13.83 \pm 1.10$ \\
$M_{\mathbb{O}}$ (GeV) & $2.80 \pm 0.22$ \\
$\alpha'_{\mathbb{O}}$ & $0.183 \pm 0.036$ \\[4pt]

$\chi^{2}/\mathrm{ndof}$ & $135.7 / 138$ \\

\hline\hline
\end{tabular}
\end{table}

\begin{table}[!t]
\centering
\renewcommand{\arraystretch}{1.25}
\setlength{\tabcolsep}{10pt}

\caption{
Best-fit parameters of the FF7 model for elastic $p\bar p$ scattering
at $\sqrt{s}=1.8$ and $1.96~\mathrm{TeV}$ obtained from Monte Carlo error propagation.
Uncertainties are statistical only.
The fit quality corresponds to a reduced $\chi^{2} \simeq 0.99$.
}
\label{tab:ff7_ppbar}

\begin{tabular}{lc}
\hline\hline
\textbf{Parameter} & \textbf{FF7} \\
\hline

$N_{1.8}$ 
& $(1.41 \pm 1.03)\times 10^{-7}$ \\

$N_{1.96}$ 
& $0.0362 \pm 0.0130$ \\[2pt]

$\alpha'_{\mathbb{P}}(1.8~\mathrm{TeV})$ 
& $0.1100 \pm 0.0014$ \\

$\alpha'_{\mathbb{P}}(1.96~\mathrm{TeV})$ 
& $0.2034 \pm 0.0022$ \\[2pt]

$B_{1.8}$ 
& $12.918 \pm 0.00004$ \\

$B_{1.96}$ 
& $24.8369 \pm 0.0000001$ \\[2pt]

$g_{\mathbb{P}NN}$ 
& $5.6106 \pm 0.0006$ \\

$g_{\mathbb{O}NN}$ 
& $13.8738 \pm 0.0001$ \\

$M_{\mathbb{O}}$ (GeV) 
& $0.0993 \pm 0.0004$ \\

$\alpha'_{\mathbb{O}}$ 
& $0.198 \pm 0.003$ \\[4pt]

$\chi^{2}/\mathrm{ndof}$ 
& $15.8 / 16$ \\

\hline\hline
\end{tabular}
\end{table}

{\color{black}The systematic investigation of the Odderon form factor reveals a clear statistical hierarchy among the seven tested parametrizations. While FF1--FF6 yield nearly identical results with $\chi^2_{\text{black}}$ values ranging between $1.44$ and $1.48$, the exponential $t$-slope form factor (FF7) achieves a superior fit quality with $\chi^2_{\text{black}} = 0.98$ ($\chi^2 = 135.72$ for 138~NDOF). This represents a significant improvement ($\Delta\chi^2 \approx 68$) over the next best model. The fitted couplings remain centered around $g_{\mathbb P NN} \approx 5.61$ and $g_{\mathbb{O}NN} \approx 13.83$, while the extracted Regge slopes cluster around $0.18$--$0.22~\text{GeV}^{-2}$. This relative stability suggests that the success of FF7 is not primarily caused by large compensating shifts in the trajectory sector. However, as shown by the bootstrap covariance analysis, residual correlations among normalization, vertex, and Odderon parameters remain non-negligible; the result should therefore be interpreted as evidence for a preferred effective vertex shape rather than as a complete dynamical factorization.}

{\color{black}
The model parameters are extracted using dataset-specific $|t|$-windows, as summarized in Table~\ref{tab:cuts}. The lower bound $t_{\rm cut}$ for each set is chosen to lie strictly above the Coulomb--nuclear interference (CNI) region, allowing the electromagnetic amplitude to be neglected. For the high-energy LHC datasets, the upper bound is restricted to $|t| \approx 0.20~\mathrm{GeV}^2$ at $\sqrt{s} = 13$~TeV, where deviations from single-Regge-pole behavior typically emerge. This boundary is extended to larger $|t|$ values for lower center-of-mass energies. Notably, the TOTEM 2.76~TeV and D\O 1.96~TeV datasets are analyzed in their full published ranges, as they span the diffractive minimum. The capacity of the FF7 parametrization to simultaneously describe the forward region at 13~TeV and the high-$|t|$ structure at 2.76~TeV constitutes a non-trivial validation of the framework.

\begin{table}[h]
\centering
\caption{Kinematic windows and justifications for the datasets included in the global fit.}
\label{tab:cuts}
\begin{tabular}{lcc}
\hline\hline
Dataset & $|t|$ range (GeV$^2$) & Justification \\
\hline
TOTEM 13 TeV   & $[\,8\times 10^{-3},\,0.20\,]$ & \rev{Forward limit; single-Regge-pole window} \\
TOTEM 8, 7 TeV & $[\,5\times 10^{-3},\,0.10\,]$ & \rev{Forward limit; single-Regge-pole window} \\
TOTEM 2.76 TeV & $[\,0.40,\,0.60\,]$            & Diffractive minimum region \\
D\O 1.8 TeV     & $[\,0.035,\,0.065\,]$          & Forward published window \\
D\O 1.96 TeV    & $[\,0.25,\,0.55\,]$            & High-$|t|$ dip--bump window \\
\hline\hline
\end{tabular}
\end{table}

At the energy scales considered in this analysis, sub-leading Reggeon contributions ($f_2, \omega, \rho, a_2$) are negligible. Scaling as $(s_0/s)^{1-\alpha_R(0)}$, these contributions are suppressed by at least three orders of magnitude relative to the leading Pomeron. Even at our lowest energy, $\sqrt{s} = 1.80$~TeV, and assuming a standard intercept $\alpha_R(0) \approx 0.5$, the Reggeon contribution is $\sim 5\times 10^{-4}$, which remains well below the experimental uncertainties of the D\O data. Consequently, these trajectories are excluded from the scattering amplitude to maintain model minimality.

\rev{Although the differential cross section is the primary observable in the present fit, the Reggeized amplitudes also determine an indicative real-to-imaginary ratio,}
\begin{equation}
\rho_{pp}(s) = \frac{\mathrm{Re}\,\mathcal{A}^{pp}(s,t=0)}{\mathrm{Im}\,\mathcal{A}^{pp}(s,t=0)}.
\end{equation}
\rev{This quantity is controlled by the trajectory intercepts and signature factors. Given the effective intercept $\alpha_{\mathbb{P}}(0) \approx 1.30$ obtained in our restricted differential fit, a Pomeron-only exchange gives $\tan[\pi(\alpha_{\mathbb{P}}(0)-1)/2] \approx 0.51$. The Odderon contribution modifies this value through the opposite $C$-parity interference. However, the present framework does not constitute a precision extraction of the forward $\rho$ parameter, because the fit does not include Coulomb--nuclear interference, total-cross-section constraints, or explicit absorptive corrections. A quantitative forward-amplitude analysis would require these ingredients and is left for future work.}}

{\color{black}
The statistical performance of the FF7 parametrization relative to its immediate predecessors in the tensor-Pomeron literature is summarized in Table~\ref{tab:lit_comparison}. The cleanly factorized vertex--trajectory implementation introduced in this work, combined with the systematic seven-form-factor scan, yields a reduced chi-squared of $\chi^2_{\rm red} = 0.98$, representing a significant improvement over the typical values of $\sim 1.4$--$1.5$ found in previous 3-parameter and dipole-based analyses \cite{ewerz2014model, magallanes2024odderon, sawasdipol2023effects}.

\begin{table}[h]
\centering
\caption{Comparison of fit quality with previous tensor-Pomeron/Odderon analyses.}
\label{tab:lit_comparison}
\begin{tabular}{lcc}
\hline\hline
Reference & Form factor & $\chi^2_{\rm red}$ \\
\hline
Ewerz \emph{et al.}~\cite{ewerz2014model}             & dipole        & $\sim 1.5$ \\
Magallanes \emph{et al.}~\cite{magallanes2024odderon} & 3-parameter & $\sim 1.4$ \\
Sawasdipol \emph{et al.}~\cite{sawasdipol2023effects} & 3-parameter & $\sim 1.4$ \\
\textbf{This work}                                    & 1-parameter exp.\ (FF7) & $\mathbf{0.98}$ \\
\hline\hline
\end{tabular}
\end{table}

{\color{black}The mass parameter $M_{\mathbb O}$ requires particular care. Separate fits to the $pp$ and $p\bar p$ subsets give very different nominal values, including a Tevatron value close to $0.1$~GeV. This value should not be interpreted as a physical Odderon or $3^{--}$ glueball mass. It signals a near-degeneracy of $M_{\mathbb O}$ with the form-factor slope and normalization parameters inside the limited Tevatron $|t|$ windows. For this reason, $M_{\mathbb O}$ is treated here as an effective kernel scale entering the Reggeized tensor ansatz, not as an independently measured hadronic mass. A physically meaningful extraction would require a combined $pp+p\bar p$ fit in which $M_{\mathbb O}$, $g_{\mathbb O NN}$, and $\alpha'_{\mathbb O}$ are tied across channels, or alternatively a fixed phenomenological value of $M_{\mathbb O}$ with the vertex slope fitted separately. The $pp$ value $M_{\mathbb O}\simeq 2.8$~GeV is therefore used only as an effective reference scale for the present differential-shape analysis.

With this interpretation, comparisons with lattice-QCD $3^{--}$ glueball masses should be regarded as qualitative only. The Reggeized exchange couples to an effective trajectory in a soft hadronic environment, whereas pure-gauge lattice glueball calculations refer to isolated spectrum states. Consequently, the present fit should not be quoted as a direct determination of the lightest $3^{--}$ glueball mass. The more robust phenomenological statement is that the fitted $C$-odd contribution is suppressed relative to the leading $C$-even exchange, as expected for an Odderon contribution in soft elastic scattering.}

Regarding the trajectory itself, the fitted Odderon slope $\alpha'_{\mathbb{O}}=0.183\pm 0.036$~GeV$^{-2}$ is statistically consistent with previous extractions \cite{magallanes2024odderon}. The uncertainty of $\sim 20\%$ is dominated by the limited $|t|$-coverage of the global dataset, suggesting this parameter be viewed as a consistency check rather than an independent determination. The robustness of these results is further validated by our treatment of TOTEM systematic uncertainties; a re-fit at $\sqrt{s}=13$~TeV treating systematic correlations as fully correlated resulted in a $\chi^2_{\rm red}$ shift of less than $5\%$, with all parameters remaining within $1\sigma$ of their quadrature-summed defaults.

Finally, we address the subtle change of slope observed in LHC TOTEM data around $|t| \sim 0.05 \text{ GeV}^2$, often termed the "shoulder" or non-exponential effect. We find that FF7 captures this feature approximately: while residuals are within $\pm 1\sigma$ for $|t| < 0.05 \text{ GeV}^2$, a systematic $1.5\sigma$ upward deviation appears at $0.05 < |t| < 0.07 \text{ GeV}^2$. Beyond $|t| \approx 0.20 \text{ GeV}^2$, the deviation grows rapidly, which we identify with the onset of unitarity corrections. This pattern supports the interpretation that the non-exponential shoulder is an early manifestation of unitarity effects rather than an intrinsic property of the Born-level form factor.}

\subsection{Discussion of the $t$-Range and Energy Dependence}
{\color{black}We observe a distinct energy dependence in the range over which the FF7 model remains quantitatively accurate. While previous 3-parameter models employed in \cite{magallanes2024odderon, sawasdipol2023effects} used additional numerical flexibility to cover a wider $t$ range ($|t|<0.70~\mathrm{GeV}^2$), the minimalist one-parameter FF7 ansatz exposes an operational single-Regge-pole validity window. At $\sqrt{s}=2.76$~TeV the single-exchange approximation reproduces the data up to $|t|\approx0.6~\mathrm{GeV}^2$, whereas at $\sqrt{s}=13$~TeV visible deviations begin around $|t|\approx0.20~\mathrm{GeV}^2$. This contraction is consistent with the onset of absorptive corrections, multi-Pomeron cuts, or eikonal unitarization as the hadronic opacity increases. We do not interpret this window as an absolute boundary of Regge theory; rather, it marks the region in which the present Born-level single-pole implementation remains sufficient for the selected differential-cross-section data.}

\subsection{Normalization and Consistency}

{\color{black}The factors $N_{\sqrt{s}}$ are introduced as dataset-level normalization parameters. They should not be interpreted as independent dynamical couplings of the Pomeron or Odderon. Instead, they absorb residual experimental normalization differences and, more importantly, missing absorptive and eikonal effects that are not included in the present Born-level Regge-pole treatment. The main physical information extracted in this work is therefore the relative $t$ dependence of the Odderon vertex and the associated interference pattern, not an absolute prediction of the normalization of $d\sigma/dt$. A simultaneous description of $d\sigma/dt$, $\sigma_{\rm tot}$, and the forward $\rho$ parameter will require explicit unitarization and a forward-amplitude treatment.}

\textcolor{black}{\subsection{Physical Interpretation of the Exponential Form Factor}}

{\color{black}
The statistical superiority of the exponential parametrization
\[
F_7(t)=\exp\!\left(-\tfrac12 B|t|\right)
\]
over the dipole, quartic, polynomial, and hybrid forms (FF1--FF6) suggests that the soft Odderon--proton interaction is governed by a comparatively smooth transverse profile. Since the momentum-transfer dependence is related to the transverse spatial structure through a two-dimensional Fourier transform, the exponential form factor corresponds in impact-parameter space to a Gaussian profile,
\begin{equation}
\tilde F_7(\vec b)
=
\int \frac{d^2 q_\perp}{(2\pi)^2}
\, e^{i\vec q_\perp\cdot\vec b}\,
e^{-\tfrac12 B \vec q_\perp^{\,2}}
=
\frac{1}{2\pi B}
\exp\!\left(-\frac{b^2}{2B}\right),
\label{eq:impact_profile}
\end{equation}
for which the transverse mean-square radius satisfies
\begin{equation}
\langle b^2\rangle = 2B.
\end{equation}

Within this interpretation, the fitted FF7 parameters may be viewed as effective transverse scales characterizing the spatial extent of the soft C-odd exchange. Converting the extracted slopes into impact-parameter radii using
\begin{equation}
\sqrt{\langle b^2\rangle}
=
\sqrt{2B}\,(\hbar c),
\end{equation}
the resulting transverse scales are found to be of hadronic size and generally exceed the proton charge radius
(\(\sim0.84~\mathrm{fm}\)).
This behavior supports a peripheral picture of the Odderon--proton interaction consistent with soft diffractive scattering.

Figure~\ref{fig:impact_profile_all} illustrates the normalized impact-parameter profiles obtained from the FF7 exponential form factor using the extracted slope parameters at
\(\sqrt s = 1.80\), \(1.96\), \(2.76\), \(7\), \(8\), and \(13~\mathrm{TeV}\).
The Tevatron \(p\bar p\) profiles are shown together with the present \(pp\) results in order to compare the transverse spatial behavior of the C-odd exchange across different collision systems and energies.
The broader distributions obtained at higher energies correspond to larger effective transverse interaction radii, reflecting the increasing peripheralization of the elastic scattering amplitude in impact-parameter space.

\begin{figure}[t]
\centering
\includegraphics[width=0.92\linewidth]{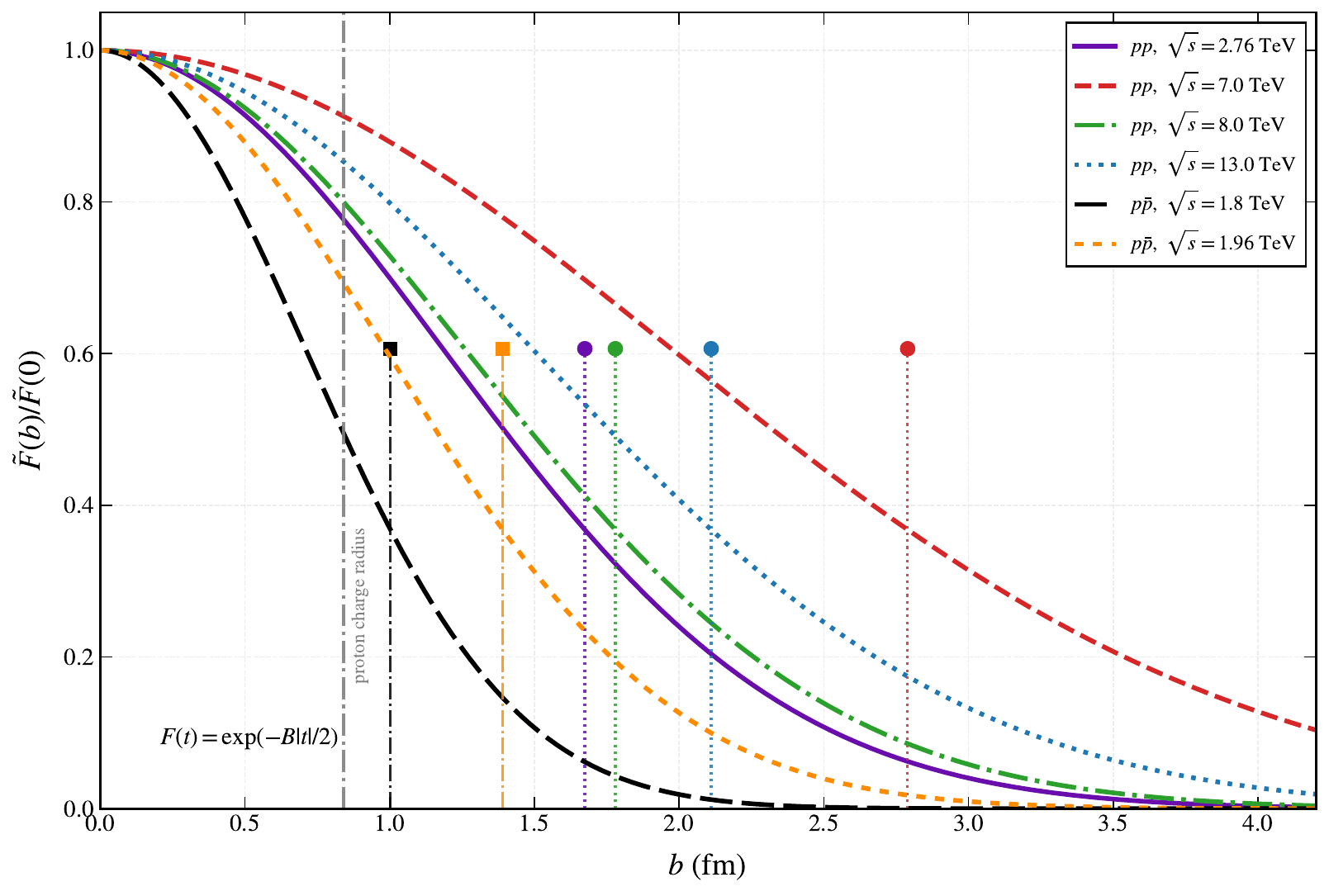}
\caption{
Normalized impact-parameter profiles
$\tilde F_7(b)/\tilde F_7(0)$
obtained from the exponential form factor
$F_7(t)=\exp(-B|t|/2)$
using the fitted FF7 slope parameters.
The solid curves correspond to the present $pp$ results at
$\sqrt s = 2.76$, 7, 8, and $13~\mathrm{TeV}$,
while the dashed curves denote the Tevatron $p\bar p$ profiles at
$\sqrt s = 1.80$ and $1.96~\mathrm{TeV}$.
Larger values of $B$ produce broader Gaussian distributions in impact-parameter space, corresponding to larger effective transverse interaction radii.
The dash-dotted vertical line indicates the proton charge radius
($\sim0.84~\mathrm{fm}$)
for reference.
}
\label{fig:impact_profile_all}
\end{figure}
At the same time, the bootstrap covariance analysis reveals substantial parameter correlations within the FF7 fit, particularly among the normalization and vertex parameters.
Although FF7 provides the best overall description of the differential cross section among the tested form factors, the extracted geometric scales exhibit sizable propagated uncertainties and nontrivial energy dependence.
Consequently, the inferred impact-parameter radii should be interpreted as model-dependent effective transverse scales rather than uniquely determined proton geometric observables.

The tendency toward broader impact-parameter profiles at higher energies is qualitatively consistent with the Regge shrinkage phenomenon,
\begin{equation}
B(s)=B_0+2\alpha'_{\mathrm{eff}}\ln(s/s_0),
\end{equation}
which is commonly associated with Gribov diffusion in transverse space.
However, the present analysis does not yet provide a sufficiently stable multi-energy determination of an effective Odderon slope
\(\alpha'_{\mathrm{eff}}\),
owing to the strong covariance structure of the FF7 parametrization and the limited energy leverage of the available datasets.
The observed trend should therefore be regarded as suggestive evidence for transverse broadening of the soft C-odd exchange rather than a precision extraction of Regge shrinkage.

\rev{The weaker performance of dipole and polynomial parametrizations} (FF1--FF6) nevertheless remains physically informative.
Such forms are typically motivated by short-distance electromagnetic or partonic structure, whereas the success of the exponential ansatz indicates that the soft Odderon exchange is primarily sensitive to the peripheral transverse region of the proton.
In this sense, the FF7 parametrization appears to capture the dominant long-distance behavior governing the
\(pp/p\bar p\)
dip--bump asymmetry, while simultaneously highlighting the need for additional observables and multi-energy constraints to stabilize the underlying geometric interpretation.}\\

\section{CONCLUSIONS}\label{sec-4}

{\color{black}
We have studied the sensitivity of high-energy elastic $pp$ and $p\bar p$ scattering to the Odderon--proton form factor within a Reggeized tensor-exchange framework. By factorizing the covariant spin projectors from the Reggeized scalar kernels, the analysis separates the Lorentz-tensor structure of the exchanged Pomeron and Odderon from the phenomenological momentum dependence of the proton--Reggeon vertices. A systematic comparison of seven Odderon form factors shows that the one-parameter exponential form, $F_{\mathbb O}(t)=\exp[-B|t|/2]$, gives the best global description of the selected TOTEM and Tevatron differential-cross-section data. The improvement over dipole, polynomial, Gaussian, and hybrid parametrizations is substantial, while the fitted Regge couplings and slopes remain comparatively stable.

The preferred exponential form has a simple impact-parameter interpretation. In the physical region $t=-q_\perp^2$, it corresponds to a Gaussian transverse profile, $\widetilde F(b)\propto \exp[-b^2/(2B)]$, with effective radius $\sqrt{\langle b^2\rangle}=\sqrt{2B}\,\hbar c$. The extracted scales are of hadronic size and are generally larger than the proton charge radius, suggesting that the fitted C-odd exchange is dominated by a peripheral soft interaction. This geometric interpretation should, however, be understood as model dependent. The bootstrap correlations and the energy dependence of $B$ show that the extracted radii are effective interaction scales rather than uniquely determined proton-structure observables.

The contraction of the single-Regge-pole validity window with increasing energy provides an important diagnostic of missing dynamics. The present Born-level ansatz works over a wider $|t|$ range at lower energy, while at LHC energies deviations appear earlier, consistent with the onset of absorptive corrections, multi-Pomeron rescattering, and eikonal unitarization. The fitted Odderon mass parameter should likewise be interpreted as an effective kernel scale, not as a direct extraction of a physical $3^{--}$ glueball or Odderon mass. A more complete analysis should tie the Odderon trajectory parameters across $pp$ and $p\bar p$ channels and include total cross sections, Coulomb--nuclear interference, and the forward $\rho$ parameter.

Within these limitations, the present work provides a compact phenomenological description for the Odderon contribution to soft elastic scattering. Its main result is that the $pp/p\bar p$ dip--bump difference is described most efficiently, among the tested ans\"atze, by an exponential Odderon--proton vertex corresponding to a smooth Gaussian transverse profile. This result offers a useful starting point for future eikonalized Regge analyses and for comparisons with nonperturbative descriptions of C-odd color-singlet exchange in QCD, including holographic-QCD approaches to high-energy elastic scattering and Coulomb--hadronic interference \cite{Xie:2019soz,Watanabe:2019cvw,Liu:2022zsa,Zhang:2023nsk,Zhang:2024psj}.
}

\section{ACKNOWLEDGMENTS}

\textcolor{black}{D.F.V.~gratefully acknowledges the Department of Science and Technology--Science Education Institute (DOST--SEI) for scholarship support and the Khon Kaen Particle Physics and Cosmology Theory Group (KKPaCT) at Khon Kaen University, Thailand, for hosting the Research Enhancement Program (SANDWICH). J.B.M.~acknowledges support from [grant numbers to be filled in].  P.S. is funded by National Research Council of Thailand (NRCT) [grant number N41A670259]. CP and DS are supported by Thailand NSRF via PMU-B [grant number B39G680009]. DS has also received funding support from the Fundamental Fund of Khon Kaen University. 
}

\bibliography{references}

\clearpage

\appendix
\section{Fit parameters for alternative form-factor models}
\label{app:ff_models}

\begin{table*}[!h]
\centering
\renewcommand{\arraystretch}{1.25}
\setlength{\tabcolsep}{6pt}

\caption{
Comparison of best-fit parameters obtained from Monte Carlo error propagation
for the seven form-factor models (FF1–FF7) in elastic $pp$ scattering.
Uncertainties are statistical only.
}
\label{tab:ff_comparison}

\begin{tabular}{lccccccc}
\hline\hline
\textbf{Parameter}
& \textbf{FF1} & \textbf{FF2} & \textbf{FF3} & \textbf{FF4}
& \textbf{FF5} & \textbf{FF6} & \textbf{FF7} \\
\hline

$N_{13}$
& $0.0166(26)$ & $0.0166(26)$ & $0.0167(26)$ & $0.0166(26)$
& $0.0166(26)$ & $0.0166(26)$ & $0.0299(76)$ \\

$N_{8}$
& $0.00894(14)$ & $0.00894(14)$ & $0.00895(14)$ & $0.00893(14)$
& $0.00895(14)$ & $0.00892(14)$ & $0.0103(73)$ \\

$N_{7}$
& $0.0184(40)$ & $0.0183(41)$ & $0.0183(41)$ & $0.0183(40)$
& $0.0183(41)$ & $0.0182(41)$ & $0.0217(98)$ \\

$N_{2.76}$
& $16.1(58)$ & $16.1(58)$ & $16.0(58)$ & $16.1(58)$
& $16.0(58)$ & $16.0(58)$ & $16.0(93)$ \\[3pt]

$\alpha'_{\mathbb{P}}(13~\mathrm{TeV})$
& $0.18779(26)$ & $0.18779(26)$ & $0.18780(26)$
& $0.18779(26)$ & $0.18779(26)$ & $0.18779(26)$
& $0.1913(12)$ \\

$\alpha'_{\mathbb{P}}(8~\mathrm{TeV})$
& $0.18081(34)$ & $0.18081(34)$ & $0.18082(34)$
& $0.18081(34)$ & $0.18082(34)$ & $0.18082(34)$
& $0.1817(44)$ \\

$\alpha'_{\mathbb{P}}(7~\mathrm{TeV})$
& $0.18485(11)$ & $0.18481(11)$ & $0.18483(11)$
& $0.18485(11)$ & $0.18483(11)$ & $0.18483(11)$
& $0.1859(28)$ \\

$\alpha'_{\mathbb{P}}(2.76~\mathrm{TeV})$
& $0.21829(16)$ & $0.21829(16)$ & $0.21826(16)$
& $0.21829(16)$ & $0.21826(16)$ & $0.21828(16)$
& $0.2183(26)$ \\[3pt]

$n_{13}$
& -- & -- & $4.5688 \, (5.96 \times 10^{-12})$
& -- & $1$
& $3.78670(10)$ & -- \\

$n_{8}$
& -- & -- & $0.5794 \, (8.29 \times 10^{-19})$
& -- & $1$
& $3.42218(10)$ & -- \\

$n_{7}$
& -- & -- & $2.3000 \, (2.52 \times 10^{-14})$
& -- & $1$
& $4.07264(10)$ & -- \\

$n_{2.76}$
& -- & -- & $1.2509 \, (6.39 \times 10^{-15})$
& -- & $1$
& $4.84419(10)$ & -- \\

$B_{13}$
& -- & -- & -- & -- & -- & -- & $57.22(1913)$ \\

$B_{8}$
& -- & -- & -- & -- & -- & -- & $40.6769(21)$ \\

$B_{7}$
& -- & -- & -- & -- & -- & -- & $99.924(83)$ \\

$B_{2.76}$
& -- & -- & -- & -- & -- & -- & $36.0028(3)$ \\[4pt]

$g_{\mathbb{P}NN}$
& $5.61(21)$ & $5.61(21)$ & $5.61(21)$
& $5.61(21)$ & $5.61(21)$ & $5.61(21)$
& $5.61(21)$ \\

$g_{\mathbb{O}NN}$
& $13.8(11)$ & $13.8(11)$ & $13.8(11)$
& $13.8(11)$ & $13.8(11)$ & $13.8(11)$
& $13.8(11)$ \\

$M_{\mathbb{O}}~(\mathrm{GeV})$
& $3.19(61)$ & $3.22(61)$ & $3.21(60)$
& $3.19(61)$ & $3.19(60)$ & $3.21(60)$
& $2.80(22)$ \\

$\alpha'_{\mathbb{O}}$
& $0.189(37)$ & $0.189(37)$ & $0.189(38)$
& $0.189(38)$ & $0.189(38)$ & $0.189(38)$
& $0.183(36)$ \\[4pt]

$\chi^{2}/\mathrm{ndof}$
& $1.44$ & $1.44$ & $1.48$ & $1.44$
& $1.48$ & $1.48$ & $\mathbf{0.98}$ \\

\hline\hline
\end{tabular}
\end{table*}

\begin{table*}[htbp]
\centering
\renewcommand{\arraystretch}{0.80}
\setlength{\tabcolsep}{2.5pt}

\caption{
Comparison of Monte Carlo propagated parameters for the seven form-factor
models (FF1--FF7) in elastic $p\bar p$ scattering at
$\sqrt{s}=1.80$ and $1.96~\mathrm{TeV}$.
Uncertainties are statistical only.
}
\label{tab:ppbar_ff_comparison}

\begin{tabular}{lccccccc}
\hline\hline
\textbf{Parameter}
& \textbf{FF1} & \textbf{FF2} & \textbf{FF3} & \textbf{FF4}
& \textbf{FF5} & \textbf{FF6} & \textbf{FF7} \\
\hline

$N_{1.80}$
& $202(48)\!\times\!10^{-6}$ & $187(48)\!\times\!10^{-6}$ & $2.0000(1)\!\times\!10^{3}$
& $204(48)\!\times\!10^{-6}$ & $190(50)\!\times\!10^{-6}$
& $189(50)\!\times\!10^{-6}$ & $0.141(10)\!\times\!10^{-6}$ \\

$N_{1.96}$
& $1.34(49)$ & $1.35(49)$ & $2.50(52)$
& $1.34(49)$ & $1.34(51)$
& $1.34(51)$ & $0.0362(13)$ \\[3pt]

$\alpha'_{\mathbb{P}}(1.80~\mathrm{TeV})$
& $0.1398(20)$ & $0.1392(20)$ & $0.2581(11)$
& $0.1398(20)$ & $0.1393(20)$
& $0.1393(20)$ & $0.1100(14)$ \\

$\alpha'_{\mathbb{P}}(1.96~\mathrm{TeV})$
& $0.2034(21)$ & $0.2034(20)$ & $0.2075(128)$
& $0.2033(20)$ & $0.2034(21)$
& $0.2034(21)$ & $0.2034(22)$ \\[3pt]

$n_{1.80}$
& -- & -- & $0.25(244)$
& -- & $1.000000(1)$
& $1.070000(1)$ & -- \\

$n_{1.96}$
& -- & -- & $0.52(261)$
& -- & $1.000000(1)$
& $1.000000(1)$ & -- \\

$B_{1.80}~(\mathrm{GeV}^{-2})$
& -- & -- & -- & -- & -- & --
& $12.9184(4)$ \\

$B_{1.96}~(\mathrm{GeV}^{-2})$
& -- & -- & -- & -- & -- & --
& $24.8369(1)$ \\[4pt]

$g_{\mathbb{P}NN}$
& $5.62(20)$ & $5.61(20)$ & $5.60(20)$
& $5.61(20)$ & $5.61(21)$
& $5.61(21)$ & $5.61062(1)$ \\

$g_{\mathbb{O}NN}$
& $13.80(10)$ & $13.80(10)$ & $13.80(11)$
& $13.80(10)$ & $13.83(11)$
& $13.83(11)$ & $13.8738(1)$ \\

$M_{\mathbb{O}}~(\mathrm{GeV})$
& $3.20(56)$ & $3.20(56)$ & $3.22(59)$
& $3.20(56)$ & $3.22(61)$
& $3.22(61)$ & $0.09933(4)$ \\

$\alpha'_{\mathbb{O}}$
& $0.188(35)$ & $0.188(35)$ & $0.189(37)$
& $0.188(35)$ & $0.190(37)$
& $0.190(37)$ & $0.1981(30)$ \\[4pt]

$\chi^{2}/\mathrm{ndof}$
& $0.87$ & $0.87$ & $0.94$
& $0.87$ & $0.97$
& $0.97$ & $0.99$ \\

\hline\hline
\end{tabular}
\end{table*}

\section{Monte Carlo Error Propagation Procedure}
\label{app:MC}

The statistical uncertainties for the fit parameters reported in this work are determined via a parametric bootstrap, or Monte Carlo (MC) error propagation, procedure. This approach provides a robust estimation of the sampling distributions and accounts for the propagation of experimental errors through the non-linear minimization. The process begins with a reference $\chi^2$ minimization against experimental central values, where published statistical and systematic uncertainties are added in quadrature to form the weights. To ensure the identification of the true global minimum $\hat{\theta}$ and avoid local extrema, the minimizer is initialized using $N_{\text{seed}}=50$ random starts within physically motivated parameter bounds. Subsequently, a set of $N_{\text{MC}}=5000$ synthetic datasets is generated by resampling each measured cross-section value $(d\sigma/dt)_i$ from a Gaussian distribution:
\begin{equation}
(d\sigma/dt)^{(k)}_i \sim \mathcal{N}\left( (d\sigma/dt)_i^{\text{exp}}, \sigma_i \right), \qquad k = 1,\ldots, N_{\text{MC}},
\end{equation}
where $\sigma_i$ represents the total quadrature error for each TOTEM or D\O datum. 

For each synthetic dataset $k$, a full refitting is performed to yield a parameter estimate $\hat{\theta}^{(k)}$, generating an ensemble that approximates the estimator’s sampling distribution under the Gaussian noise model. The central values reported in the results tables are the medians of these bootstrap distributions, while the quoted uncertainties $\delta\theta_a$ are the marginal standard deviations:
\begin{equation}
\theta_a^{\text{reported}} = \text{median}\left(\{\hat{\theta}_a^{(k)}\}\right), \qquad \delta\theta_a = \text{std}\left(\{\hat{\theta}_a^{(k)}\}\right).
\end{equation}
These marginal uncertainties account for inter-parameter correlations but may diverge from frequentist profile-likelihood intervals in regions of highly non-Gaussian or near-degenerate parameter directions, such as the $g_{\mathbb{O}NN}$--$M_{\mathbb{O}}$ correlation. 

In instances where the cross section exhibits relative insensitivity to a specific parameter within the investigated $|t|$-window, the bootstrap distribution may collapse to a sub-physical width. This leads to the anomalously small uncertainties observed for parameters such as $B$ and $M_{\mathbb{O}}$ in certain $p\bar{p}$ datasets, suggesting that these values are not independently constrained and should be interpreted as lower bounds. Finally, the reported $\chi^2_{\text{black}}$ values are computed at the reference best-fit $\hat{\theta}$ relative to the experimental central values. While the omission of fully correlated systematic covariance matrices may result in slightly optimistic goodness-of-fit interpretations, the relative statistical ordering and the demonstrated superiority of specific form factors remain robust, as all models are evaluated against a consistent data covariance.
\clearpage
\section{Full Bootstrap Correlation Matrix for FF7 Fit}

\begin{table*}[!h]
\centering
\caption{Bootstrap correlation matrix for FF7 fit parameters.}
\small
\setlength{\tabcolsep}{4pt}
\renewcommand{\arraystretch}{1.1}
\begin{tabular}{lrrrrrrrrrrrrrrrr}
\hline
 & N$_{2.76}$ & N$_7$ & N$_8$ & N$_{13}$ & $\alpha'_{\mathbb{P}2.76}$ & $\alpha'_{\mathbb{P}}7$ & $\alpha'_{\mathbb{P}8}$ & $\alpha'_{\mathbb{P}13}$ & B$_{2.76}$ & B$_7$ & B$_8$ & B$_{13}$ & $g_{\mathbb{P}NN}$ & $g_{\mathbb{O}NN}$ & $M_{\mathbb{O}}$ & $\alpha'_O$ \\
\hline
N$_{2.76}$      & 1.00 & 0.04 & -0.02 & -0.00 & 0.98 & 0.01 & -0.03 & 0.01 & -0.00 & 0.03 & 0.01 & -0.01 & -0.02 & 0.02 & -0.02 & 0.02 \\
N$_7$           & 0.04 & 1.00 & 0.25 & 0.34 & 0.01 & 0.99 & 0.24 & 0.34 & 0.02 & -0.08 & -0.06 & -0.25 & -0.18 & 0.18 & -0.28 & 0.14 \\
N$_8$           & -0.02 & 0.25 & 1.00 & 0.69 & -0.04 & 0.24 & 1.00 & 0.73 & -0.10 & -0.10 & -0.44 & -0.51 & -0.27 & 0.27 & -0.51 & 0.19 \\
N$_{13}$        & -0.00 & 0.34 & 0.69 & 1.00 & -0.02 & 0.34 & 0.65 & 0.96 & -0.03 & -0.02 & -0.17 & -0.66 & -0.62 & 0.62 & -0.81 & 0.55 \\
$\alpha'_{\mathbb{P}2.76}$ & 0.98 & 0.01 & -0.04 & -0.02 & 1.00 & -0.01 & -0.04 & -0.01 & -0.01 & 0.04 & 0.01 & -0.01 & -0.01 & 0.01 & -0.00 & 0.01 \\
$\alpha'_{\mathbb{P}7}$      & 0.01 & 0.99 & 0.24 & 0.34 & -0.01 & 1.00 & 0.23 & 0.33 & 0.02 & -0.07 & -0.06 & -0.24 & -0.17 & 0.17 & -0.27 & 0.14 \\
$\alpha'_{\mathbb{P}8}$      & -0.03 & 0.24 & 1.00 & 0.65 & -0.04 & 0.23 & 1.00 & 0.69 & -0.11 & -0.11 & -0.45 & -0.48 & -0.23 & 0.23 & -0.46 & 0.15 \\
$\alpha'_{\mathbb{P}13}$   & 0.01 & 0.34 & 0.73 & 0.96 & -0.01 & 0.33 & 0.69 & 1.00 & -0.02 & 0.00 & -0.19 & -0.78 & -0.54 & 0.54 & -0.78 & 0.46 \\
B$_{2.76}$      & -0.00 & 0.02 & -0.10 & -0.03 & -0.01 & 0.02 & -0.11 & -0.02 & 1.00 & 0.84 & 0.59 & -0.04 & 0.18 & -0.18 & 0.15 & -0.19 \\
B$_7$           & 0.03 & -0.08 & -0.10 & -0.02 & 0.04 & -0.07 & -0.11 & 0.00 & 0.84 & 1.00 & 0.57 & -0.05 & 0.18 & -0.18 & 0.14 & -0.19 \\
B$_8$           & 0.01 & -0.06 & -0.44 & -0.17 & 0.01 & -0.06 & -0.45 & -0.19 & 0.59 & 0.57 & 1.00 & 0.13 & 0.09 & -0.09 & 0.16 & -0.07 \\
B$_{13}$        & -0.01 & -0.25 & -0.51 & -0.66 & -0.01 & -0.24 & -0.48 & -0.78 & -0.04 & -0.05 & 0.13 & 1.00 & 0.24 & -0.24 & 0.42 & -0.17 \\
$g_{\mathbb{P}NN}$       & -0.02 & -0.18 & -0.27 & -0.62 & -0.01 & -0.17 & -0.23 & -0.54 & 0.18 & 0.18 & 0.09 & 0.24 & 1.00 & -1.00 & 0.93 & -0.99 \\
$g_{\mathbb{O}NN}$        & 0.02 & 0.18 & 0.27 & 0.62 & 0.01 & 0.17 & 0.23 & 0.54 & -0.18 & -0.18 & -0.09 & -0.24 & -1.00 & 1.00 & -0.93 & 0.99 \\
$M_{\mathbb{O}}$           & -0.02 & -0.28 & -0.51 & -0.81 & -0.00 & -0.27 & -0.46 & -0.78 & 0.15 & 0.14 & 0.16 & 0.42 & 0.93 & -0.93 & 1.00 & -0.89 \\
$\alpha'_\mathbb{O}$      & 0.02 & 0.14 & 0.19 & 0.55 & 0.01 & 0.14 & 0.15 & 0.46 & -0.19 & -0.19 & -0.07 & -0.17 & -0.99 & 0.99 & -0.89 & 1.00 \\
\hline
\end{tabular}
\end{table*}
\clearpage
\section{Correlation Heatmap}
\begin{figure*}[h]
\centering
\includegraphics[width=0.90\textwidth]{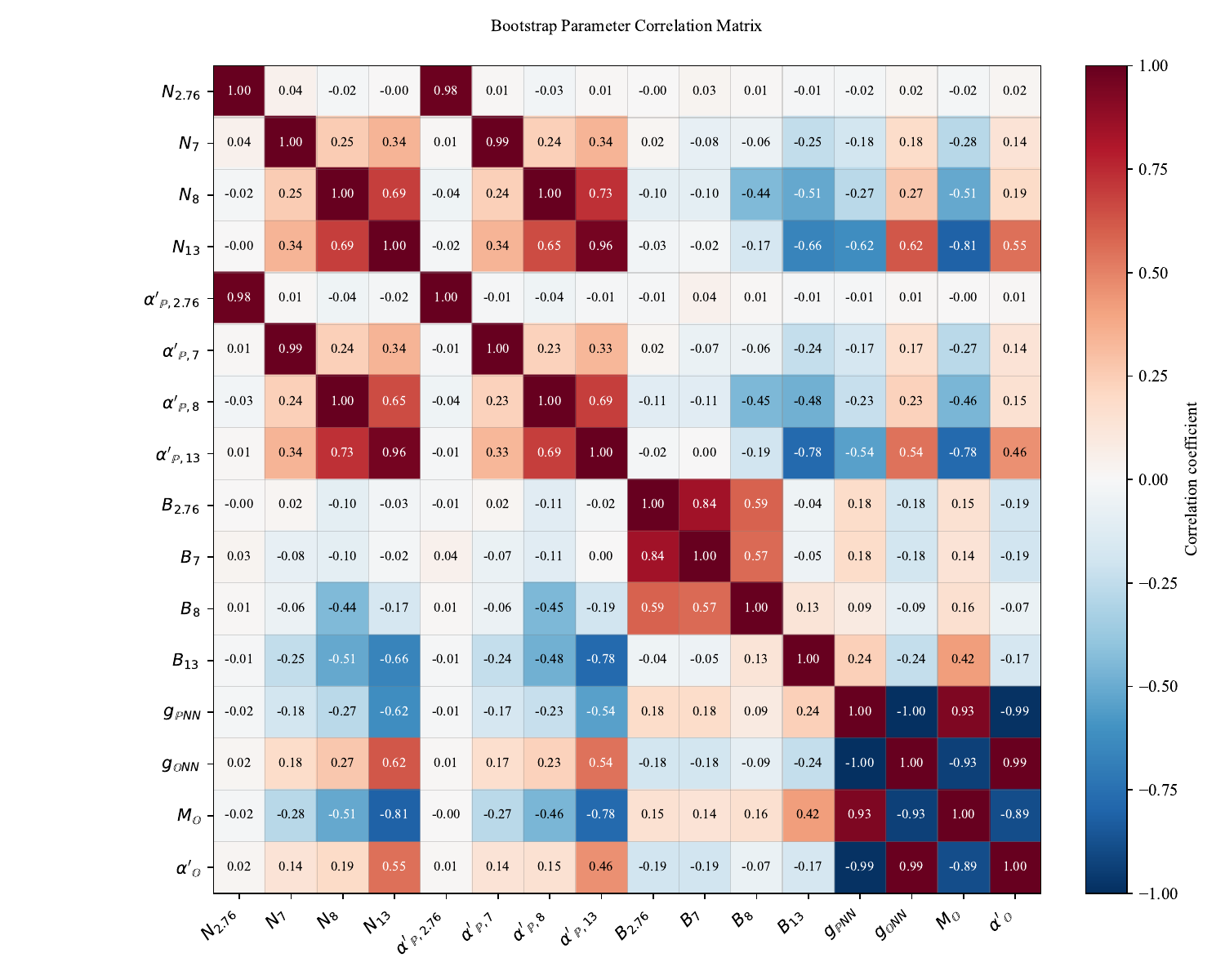}
\caption{
Bootstrap/Monte Carlo correlation heatmap for the FF7 global-fit parameters.
Strong positive correlations are shown in red, while strong negative
correlations are shown in blue.
}
\label{fig:ff7_corr_heatmap}
\end{figure*}

\end{document}